\documentclass[preprint, prfluids, aps,  longbibliography]{revtex4-2}

\usepackage{xcolor} 

\usepackage{graphicx}
\usepackage{dcolumn}
\usepackage{bm, amsmath}
\usepackage{amsfonts}
\usepackage{subcaption}
\usepackage{siunitx}
\usepackage[english]{babel}\href{}{}
\begin{document}

\title{\textbf{Characterizing periodic orbits in two-dimensional
Rayleigh-Bénard flows}}

\author{Joaquín Cullen}
\affiliation{Universidad de San Andrés, Buenos Aires, Argentina}

\author{Melisa Y. Vinograd}
\affiliation{Universidad de San Andrés, Buenos Aires, Argentina}
\affiliation{Departamento de Física, Universidad de Buenos Aires, CABA, Argentina}

\author{Patricio Clark Di Leoni}
\email{Contact author: pclarkdileoni@udesa.edu.ar}
\affiliation{Universidad de San Andrés, Buenos Aires, Argentina}
\affiliation{CONICET, Argentina}

\date{May 2026}

\newcommand{\CITE}[1]{\textcolor[HTML]{E76F51}{\textbf{[Citation needed]}}}
\newcommand{\MELISA}[1]{\textcolor[HTML]{2A9D8F}{\textbf{[Melisa: #1]}}} 
\newcommand{\PATRICIO}[1]{\textcolor[HTML]{D1495B}{\textbf{[Patricio: #1]}}} 
\newcommand{\JOAQUIN}[1]{\textcolor[HTML]{2780F5}{\textbf{[Joaquín: #1]}}} 

\newcommand{\REVA}[1]{#1}
\newcommand{\REVB}[1]{#1}

\newcommand{\Rey}{\mathrm{Re}}
\newcommand{\Ray}{\mathrm{Ra}}
\newcommand{\Prn}{\mathrm{Pr}}
\newcommand{\Nuss}{\mathrm{Nu}}
\newcommand{\fref}[1]{\figurename~\ref{#1}}

\begin{abstract}
Unstable periodic orbits and steady states are believed to form the backbone of spatiotemporal chaos and turbulence, yet their computation in thermally driven flows remains scarce for transitional regimes. In this work we compute and characterize a steady state and three families of periodic orbits in two-dimensional Rayleigh-Bénard at $\Pr=1$, near the transition to chaos. We find that in its route to chaos, the flow hops between several sets of orbits after becoming quasiperiodic. We use Floquet analysis to study the stability of the orbits obtained and characterize their bifurcations, showing how the appearance of primary and secondary frequencies, as well as phase-locking mechanisms, are all related to the dynamics of the orbits.  We study in detail how the flow shadows the orbits found and determine in which regimes each orbit is dynamically relevant or not. Our analysis also reveals two important insights: (1) all symmetries are broken before the onset of chaos, and (2) this onset does not alter the behavior of the heat transport.
\end{abstract}

\maketitle

\section{Introduction}

The route-to-chaos of many fluid flows follows by now familiar paths of steady states succeeded by periodic orbits that become quasiperiodic and that eventually lead to turbulence.
Rayleigh-Bénard flow, in both two and three dimensions, is one of the most famous examples of such behavior~\cite{dijkstra_bifurcation_2023}.
Early experimental studies~\cite{gollub_many_1980} already revealed such sequences of bifurcations. 
However, detailed analysis reveals that this picture can be far more intricate, with systems switching back between periodic and quasiperiodic solutions and non-ergodic regimes appearing, all marred with numerical and experimental effects and complications. 
While characterizations of the regimes and bifurcations of the flow can be done by direct inspection, it is the underlying orbits that hold the key to its dynamics.

Periodic orbits occupy a central place in dynamical systems theory, serving as the building blocks of chaotic attractors~\cite{auerbach_exploring_1987,lan_Cycle_2010,budanur_tutorial_2015}. In the realm of fluid mechanics, they were traditionally calculated from lower-order truncated systems. However, over the past two decades, advances in numerical methods, particularly Newton-Krylov solvers~\cite{viswanath_recurrent_2007,gibson_visualizing_2008}, have enabled the computation of invariant solutions directly from the full system of partial differential equations governing the flow. Unstable Periodic Orbits (UPOs) have since been calculated for many flows, including pipe flow~\cite{viswanath_recurrent_2007,budanur_relative_2017,duguet_relative_2008}, Couette flow~\cite{viswanath_critical_2008,gibson_visualizing_2008,kawahara_Periodic_2001,kawahara_Significance_2012}, Taylor-Couette flow~\cite{crowley_turbulence_2022}, Kolmogorov flow~\cite{Chandler_Kerswell_2013,suri_Capturing_2020}, and flow past a sphere~\cite{schuh_frantz_bifurcation_2025}, among others. 
\REVB{Closer to the configuration studied here, invariant solutions have also proven central in inclined layer convection, where the bifurcations and dynamical connections of a large set of invariant states have been mapped out \cite{reetz_invariant_2020,reetz_invariant_2020a}, as well as in doubly diffusive convection, where steady, periodic, and spatially localized states organize the dynamics near onset \cite{bergeon_spatially_2008,beaume_homoclinic_2011}.}
A recurring theme across all these works is how the key features and statistics of each case can be reconstructed from the underlying invariant solutions. That said, not all orbits are dynamically relevant and there are alternative ways to characterize flow statistics in terms of UPOs~\cite{cleary_Dynamical_2025,redfern_Dynamically_2024}. Recent methodological advances include variational methods~\cite{parker_variational_2022} and machine learning techniques~\cite{beck_machine-aided_2024,page_exact_2024}.

In the specific context of Rayleigh-Bénard (RB) convection, the study of coherent structures and bifurcation sequences has a long history. Early numerical work~\cite{curry_order_1984} mapped transitions from ordered to disordered states in both two and three-dimensional geometries. This program was extended with an analysis of symmetry breakings during bifurcations~\cite{zienicke}, a detailed bifurcation analysis reporting multistable regimes in which different initial conditions settle onto distinct attractors~\cite{paul_bifurcation_2012,paul_order_2009}, and the calculation of periodic orbits and their instabilities obtained from low-order Galerkin projections~\cite{puigjaner_steady_2011}.
More recently, attention has turned to exact invariant solutions as a framework for understanding heat transport. Families of steady convective rolls have been computed and connected to optimal heat-transport bounds~\cite{waleffe_heat_2015,sondak_optimal_2015}, and a comprehensive study of steady states across a wide range of Rayleigh numbers has been carried out~\cite{wen_steady_2022}. Coherent invariant solutions and their role in the transition to turbulence in two-dimensional RB convection have also been studied explicitly, with steady states used to explain heat transport scaling~\cite{kooloth_coherent_2021}. Despite these advances, a systematic computation and characterization of the periodic orbits underpinning the observed transitions in two-dimensional RB flows, and their relationship to the route to chaos, remains to be done.

In this work, we address that gap. We compute and track, via Newton-Krylov-Hookstep continuation from the full system of partial differential equations, a steady state and three families of periodic orbits across a broad range of Rayleigh numbers, spanning the onset of convection through the chaotic regime. We use Floquet analysis to determine the stability of each solution and to identify the bifurcation type at each change of stability. We also perform direct comparisons with trajectories calculated from direct numerical simulations to show how the invariant solutions shadow the observed dynamics, account for the frequencies present in the flow, and organize the sequences of transitions, from periodic motion through quasiperiodicity, phase-locking, and ultimately to chaos. 

The remainder of the paper is organized as follows. Section~\ref{sec:methods} describes the governing equations, the symmetries of the system, the numerical simulations, and the methods used to compute and continue the invariant solutions. Section~\ref{sec:results} presents the results: it first characterizes the flow regimes through spectral analysis (\ref{sec:regimes}), describes the invariant solutions and their Floquet stability (\ref{sec:characterization}), compares them with the observed dynamics (\ref{sec:comparison}), analyzes the route to chaos via symmetry breaking and Lyapunov exponents (\ref{sec:chaos}), and quantifies the impact of the transitions on global heat transport. Conclusions are drawn in Section~\ref{sec:conclusions}.

\section{Problem set-up and numerical methods}
\label{sec:methods}

\subsection{Governing equations}

We study Rayleigh-Bénard convection, which models the buoyancy-driven motion of a fluid confined between two horizontal plates maintained at constant temperatures. The bottom-plate is held at temperature $T_b$, while the top plate is cooler by a temperature difference $\Delta T$. The Boussinesq approximation is adopted, wherein density is assumed to vary linearly with temperature, and compressibility effects are retained only in the buoyancy term. In a rectangular domain of height $h$ and width $L$, with Cartesian coordinates $\boldsymbol{x}=(x,\,z)$, the velocity $\boldsymbol v = (u,\,w)$, pressure $p$, and temperature $T$ evolve according to

\begin{align}
\partial_t \boldsymbol v + (\boldsymbol v \cdot \nabla)\boldsymbol v
&= -\nabla p + \nu \nabla^2 \boldsymbol v + \alpha g\, T\, \hat{\boldsymbol z},\\
\nabla \cdot \boldsymbol v &= 0,\\
\partial_t T + (\boldsymbol v \cdot \nabla) T &= \kappa \nabla^2 T,
\label{gov_eq}
\end{align}
where $\nu$ is the kinematic viscosity, $\kappa$ the thermal diffusivity, $\alpha$ the thermal expansion coefficient, $g$ the gravitational acceleration, and $\hat{\boldsymbol z}$ points opposite gravity.
In the absence of motion, the system admits a conductive equilibrium state, in which the temperature varies linearly with the height. 
\begin{equation}
T_0(z) = T_b - z\, \Delta T / h.
\end{equation}
Consequently, it is convenient to work with a rescaled temperature fluctuation relative to this linear profile
\begin{equation}    
\theta = (T - T_0)\, \sqrt{\alpha g h / \Delta T},
\end{equation}
where $T_0$ is the linear profile, and $\theta$ is the temperature fluctuation field. 

The dynamics of the system are mainly dictated by two non-dimensional control parameters. The Rayleigh number,
\begin{equation}    
\Ray = \frac{\alpha g h^3 \Delta T}{\nu \kappa},
\end{equation}
measures the balance of buoyancy against diffusive (thermal and viscous) effects,
while the Prandtl number
\begin{equation}
\Prn=\nu/\kappa,
\end{equation} 
prescribes the ratio between the dissipative mechanisms. The aspect ratio $L/h$ also has an effect on the dynamics.

Several characteristic scales and quantities of the flow can be defined from the aforementioned parameters. The free-fall velocity,
\begin{equation}
    U = \sqrt{g\,h\,\alpha\, \Delta T} ,
\end{equation}
corresponds to the velocity of a fluid parcel accelerated by buoyancy over a distance $h$. The associated time scale is $t' = h/U$, and all temporal quantities reported in this work are expressed in units of $t'$. Another quantity of interest is the Nusselt number $\Nuss$,
\begin{equation}
\Nuss(z) = \frac{\langle w T - \kappa\, \partial_z T \rangle_{x,t}}{\Delta T\, \kappa / h},
\end{equation}
where $\langle\cdot \rangle_{x,t}$ denotes an average over horizontal direction and time. For statistically stationary states and under periodic lateral boundary conditions, the total vertical heat flux is conserved, and thus $\Nuss$ is independent of $z$. Values of $\Nuss>1$ indicate enhanced heat transport due to convection.





The governing equations of the system are equivariant under a discrete symmetry group that constrains the dynamics and organizes the structure of invariant solutions~\cite{zienicke}. The symmetry group is given by  
\begin{equation}  
\mathcal{G} = \{ \mathbb{I}, S_1, S_2, S_3 \},  
\label{eq:syms}  
\end{equation}  
where $\mathbb{I}$ denotes the identity. The nontrivial elements act on the spatial coordinates $(x,z)$, and fields $\boldsymbol{v}=(u,w)$, and $\theta$ as  
\begin{align}  
S_1:\quad (x,z) &\mapsto (-x, z),  
& (u,w,\theta) &\mapsto (-u, w, \theta), \label{eq:S1}\\
S_2:\quad (x,z) &\mapsto (-x + L/2, h - z),  
& (u,w,\theta) &\mapsto (-u,-w,-\theta), \label{eq:S2}\\  
S_3:\quad (x,z) &\mapsto (x + L/2, h - z),  
& (u,w,\theta) &\mapsto (u,-w,-\theta). \label{eq:S3}  
\end{align}
The transformation $S_1$ corresponds to a reflection about the vertical mid-plane, while $S_2$ combines a reflection about the horizontal mid-plane with a horizontal translation by $L/2$, admissible because of the periodic boundary conditions in $x$. 
$S_3$ then follows as the composition of $S_1$ and $S_2$.
The associated sign changes in $(u,w,\theta)$ ensure invariance of the governing equations under these operations. Individual solutions, however, need not be invariant under these symmetries and may instead form symmetry-related families.

\subsection{Numerical simulations}

Direct numerical simulations (DNS) were performed using the Special Periodic
Continuation Turbulence Solver (\texttt{SPECTER})~\cite{fontana_fourier_2020} to evolve the governing equations \eqref{gov_eq}.
The code employs a pseudospectral method with MPI-OpenMP-CUDA parallelization, and uses Fourier continuation to handle the non-periodic boundary conditions in the vertical direction. 
\REVA{In this approach, the non-periodic fields are extended over a small auxiliary region beyond the physical domain so as to construct smooth periodic extensions (using the FC-gram methodology), which allows all spatial derivatives to be computed via fast Fourier transforms. Time integration is fully explicit, carried out with a second-order Runge-Kutta scheme, and the incompressibility constraint is enforced through a pressure projection method, by solving a Poisson equation for the pressure at each Runge-Kutta stage.}
Nonlinear terms were de-aliased using the standard $2/3$ rule. 
\REVA{The time step was chosen so as to keep the advective CFL number below unity, which resulted in $\Delta t$ ranging between $\num{5e-4}$ and $10^{-3}$ free-fall time units. }

All simulations were conducted in a domain of size
$[L,\,h]=[2\pi,\,\pi]$, with aspect ratio $L/h = 2$. 
Periodic boundary conditions were imposed in $x$ and no-slip walls at the top and bottom.
The Prandtl number was fixed at $\Prn=1$, comparable to the value for air. 
A total of 34 Rayleigh numbers were considered, spanning the range $10^5<\Ray<\num{2e7}$. 
The numerical resolution $[N_x,\,N_z]$ 
was increased with $\Ray$ as: $[256,\,103],\; [256,\,231],\;[512,\,512]$ for the
respective Rayleigh ranges $[10^5,\,\num{4e6}],\;[\num{4e6},\,
\num{8e6}],\;[\num{8e6}, \num{2e7}]$, in all cases chosen to adequately resolve the smallest relevant scales. 
\REVB{Specifically, two standard criteria were verified a posteriori for every run. First, the maximum resolved (de-aliased) wavenumber satisfies $k_{\max}\eta \gtrsim 1.5$ at all times, where $\eta = (\nu^3/\epsilon )^{1/4}$ is the Kolmogorov scale computed from the instantaneous volume averaged dissipation rate, above the usual threshold value of $k_{\max}\eta \gtrsim 1$ for spectral DNS~\cite{grotzbach_spatial_1983,pope_turbulent_2000}. 
Second, the thermal boundary layers remain covered by 4-10 grid points at all $\Ray$, satisfying the criterion~\cite{shishkina_boundary_2010}, which for our parameters require a minimum of 3 to 4 grid points. In addition, runs at boundaries of resolution ranges were repeated at both resolutions, yielding Nusselt numbers that agree to within 1\%.
} 

\subsection{Computation of periodic orbits and steady states}

Let $\boldsymbol{X}\in \mathbb{R}^n$ denote the state vector of the system, comprising the velocity and temperature fields at every grid point in a one-dimensional arrangement, and let $\Phi^t$ denote the flow-map that advances $\boldsymbol{X}$ forward in time by $t$ according to the governing equations. Exact invariant solutions are defined as fixed points of $\Phi^t$.
A steady state $\boldsymbol{X}_0$ satisfies $\Phi^t(\boldsymbol{X}_0)=\boldsymbol{X}_0$ for all $t$, while a periodic orbit of period $\tau$ satisfies $\Phi^\tau(\boldsymbol{X}_0) = \boldsymbol{X}_0$. Because the equations are invariant under continuous horizontal translation due to the periodic boundary conditions, relative periodic orbits can also arise, satisfying
\begin{equation}
\mathcal{T}_{s}\Phi^\tau(\boldsymbol{X}_0) = \boldsymbol{X}_0,
\end{equation}
where $\mathcal{T}_{s}$ denotes a horizontal shift by $s$. 

To converge the invariant states we recast these equations as nonlinear root-finding problems. 
Collecting the unknowns into $\Tilde{\boldsymbol{X}} = [\boldsymbol{X},\tau, s]^T$, the RPO condition becomes
\begin{equation}
\boldsymbol{F}(\Tilde{\boldsymbol{X}}) 
\equiv 
\mathcal{T}_{s}\Phi^\tau(\boldsymbol{X}) - \boldsymbol{X} = \boldsymbol{0}.
\end{equation}
Analogous equations can be defined for steady states and periodic orbits.
Candidate initial guesses for the Newton solver were obtained from a recurrence analysis of the DNS trajectories via the function
\begin{equation}
    G(t, \tau) = \underset{s}{\text{min}} \frac{\Vert \mathcal{T}_{s} \boldsymbol{X}(t+\tau) - \boldsymbol{X}(t) \Vert}{\Vert \boldsymbol{X}(t)\Vert},
\end{equation}
where $\Vert \cdot \Vert$ denotes the $L_2$ norm. Local minima of $G$ identify near-recurrences $(t,\tau,s)$ which are then supplied as starting points for the iterative method. 
\REVB{Note that, as it operates on DNS data, this recurrence analysis is naturally biased towards dynamically relevant solutions. It detects weakly unstable ones, which the trajectory transiently shadows, but may miss strongly unstable ones. The continuation procedure described below mitigates this bias by tracking converged solutions into parameter ranges where the DNS no longer approaches them.}

We apply an adaptation of the Newton-GMRes-Hookstep method of Viswanath~\cite{viswanath_critical_2008}, implemented in the open-source Python library \texttt{spookyflows}~\cite{spookyflows_2025}, to solve these non-linear equations. In-depth descriptions of the general method can be found in~\cite{Chandler_Kerswell_2013,willis_equilibria_2019}.
In the RPO case, the Newton iteration takes the form
\begin{equation}
\left( \mathcal{T}_{s_i} D\Phi^{\tau_i}(\boldsymbol{X}_i)-\mathbb{I}\right)\, \delta \boldsymbol{X} + 
\mathcal{T}_{s_i}\frac{  \partial \Phi^{\tau_i}(\boldsymbol{X}_i)}{\partial t}\, \delta \tau+
\mathcal{T}_{s_i} \frac{ \partial \Phi^{\tau_i}(\boldsymbol{X}_i)}{\partial x}\,\delta s 
= -\boldsymbol{F}(\Tilde{\boldsymbol{X}_i}),
\label{newtonlin}
\end{equation}
where $i$ denotes the Newton iteration, $[\delta \boldsymbol{X}, \delta \tau, \delta s]^T = \Tilde{\boldsymbol{X}}_{i+1} - \Tilde{\boldsymbol{X}_{i}}\,$ is the Newton step, and $D\Phi^\tau(\boldsymbol{X})$ is the Jacobian of the time-$\tau$ flow map with respect to the state vector. 
However, in high-dimensional discretizations the Jacobian is never formed explicitly. Instead, we employ matrix–vector products approximated by finite differences, i.e.
\begin{equation}
 D\Phi^\tau(\boldsymbol{X}_i)\;
 \delta \boldsymbol{X}
 \; \approx \;
\frac{\Phi^\tau(\boldsymbol{X}_i+\epsilon \delta \boldsymbol{X}) - \Phi^\tau(\boldsymbol{X}_i)}
{\epsilon}\; , \label{eq:findiff}
\end{equation}
with $\epsilon$ such that $\epsilon\Vert \delta \boldsymbol{X}\Vert = \sqrt{\epsilon_{\mathrm{mach}}}\Vert \Tilde{\boldsymbol{X}_i}\Vert$, where $\epsilon_\mathrm{mach}$ is the machine tolerance, given that it balances truncation against round-off error. GMRes~\cite{trefethen_numerical_1997} is then used to iteratively solve \eqref{newtonlin} without ever constructing or storing the Jacobian explicitly.

Furthermore, since there are two more unknowns than equations (given by $\tau$ and $s$), additional equations are imposed which eliminate the degeneracies associated with the system's invariance to translation in time and the horizontal direction:
\begin{equation}
    \delta \boldsymbol{X}_i \cdot \frac{\partial \boldsymbol{X}_i}{\partial x} = \delta \boldsymbol{X}_i \cdot \frac{\partial \boldsymbol{X}_i}{\partial t} = 0.
\end{equation}
Each evaluation of the flow-map $\Phi^t$ is carried out by direct calls from the Newton solver to the DNS code.

Newton iterations may fail to converge if the initial guess is not close to a true solution. To stabilize the process, we employ a Hookstep trust-region strategy. Instead of taking the full Newton step $\delta \Tilde{\boldsymbol{X}}$, we solve a constrained minimization problem in which the residual norm $\Vert \boldsymbol{F}(\Tilde{\boldsymbol{X}}+\delta \Tilde{\boldsymbol{X}})\Vert$ is reduced subject to $\Vert \delta \Tilde{\boldsymbol{X}} \Vert \leq \Delta$, where $\Delta$ is an adaptively chosen radius. If the step decreases the residual, the trust region is expanded, otherwise, it is contracted.

In our implementation, convergence was declared once the relative residual dropped below
\begin{equation}
\frac{\Vert \boldsymbol{F}(\tilde{\boldsymbol{X}}) \Vert} {\Vert \boldsymbol{X} \Vert} \leq 10^{-8}.
\end{equation}
Typically, between $5$ and $20$ Newton iterations were required for convergence of invariant solutions. Each Newton step required $10$–$30$ GMRes iterations, depending on the Rayleigh number, with each GMRes iteration involving one evaluation of the flow map difference. Consequently, the computational cost of converging a single invariant solution is equivalent to a few hundred DNS time steps of length $\tau$.

Because the GMRes method is matrix-free, the memory requirements scale essentially linearly with the size of the state vector, i.e. the number of velocity and temperature degrees of freedom. For the resolutions considered here, 
\REVB{the storage of the Krylov basis, which consisted of one vector of the size of the full state vector per GMRes iteration (i.e., between 10 and 30 vectors), dominated the memory footprint}
while the primary cost arose from repeated evaluations of $\Phi^\tau$, which were fully parallelized in \texttt{SPECTER}. In practice, the Hookstep mechanism successfully reduced stagnation and ensured monotonic decrease of the residual.



Once a periodic orbit or steady state was converged, we track its evolution across varying Rayleigh numbers using continuation methods. We primarily employ natural parameter continuation, utilizing a converged solution at a given $\Ray$ as an initial guess for an incrementally shifted $\Ray$. 
We use pseudo-arclength continuation in cases where the natural parameter continuation failed to converge~\cite{Chandler_Kerswell_2013}.

To assess the linear stability of the periodic orbits, we calculate the Floquet multipliers and modes ($\mu\in \mathbb{C}$ and $\boldsymbol{\xi}\in \mathbb{C}^n$  respectively). These relate to the eigenvalues and eigenvectors of the monodromy matrix $D\Phi^\tau(\boldsymbol{X})$, i.e. the Jacobian of the time-$\tau$ flow map evaluated at a point on the periodic orbit of period $\tau$.  Given the finite-difference approach introduced in Eq.~\eqref{eq:findiff}, both can be estimated via the Arnoldi method~\cite{viswanath_recurrent_2007}. For each converged state we computed 100 eigenvalues and eigenvectors, and since the Arnoldi iteration converges fastest for the extremal eigenvalues~\cite{trefethen_numerical_1997}, the potentially unstable directions are well resolved.   


\section{Results}
\label{sec:results}

\subsection{Flow regimes and transitions}
\label{sec:regimes}

\begin{figure}
    \centering
    \includegraphics[width=0.9\linewidth]{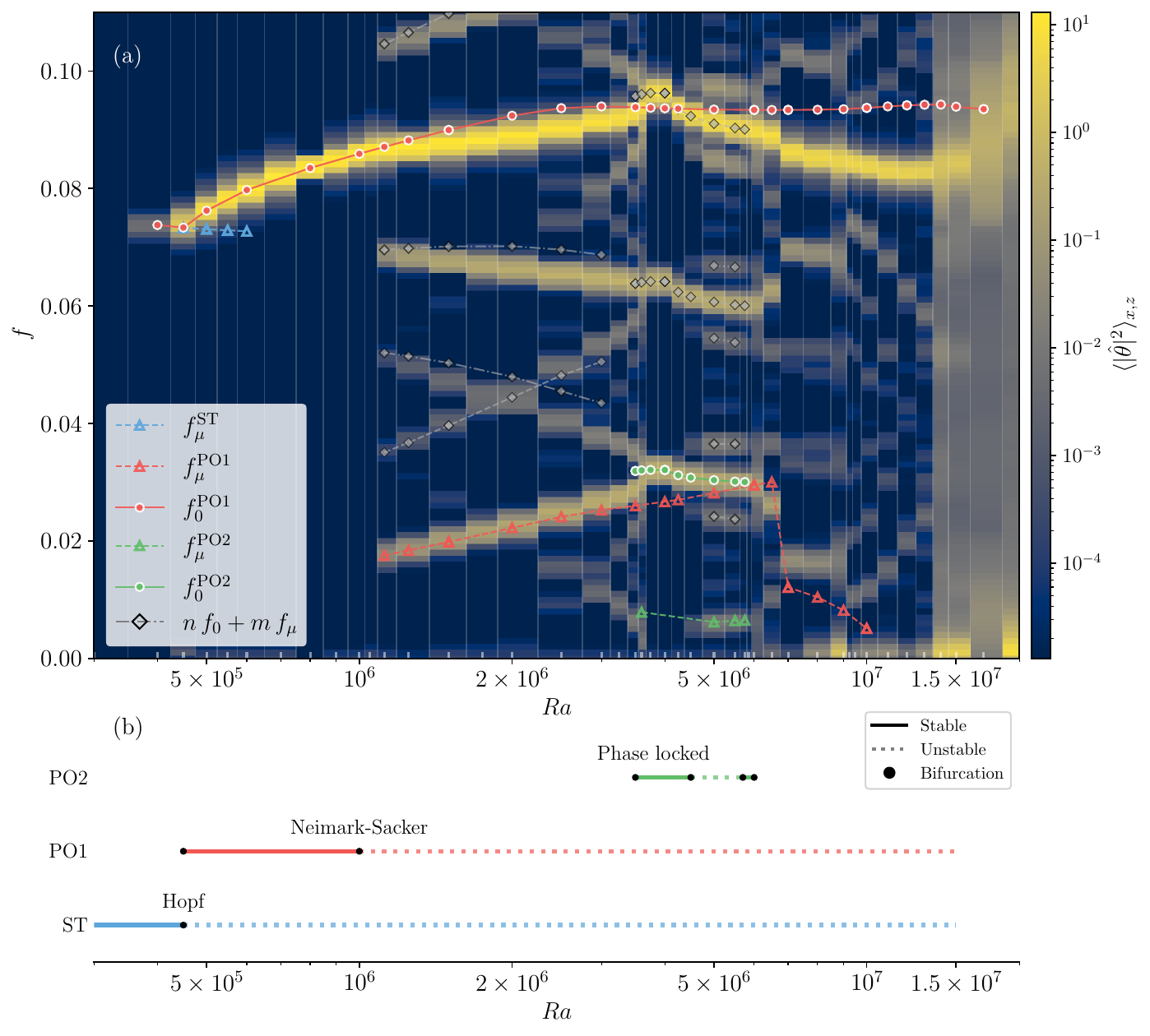}    
    \caption{\REVA{Overview of the flow regimes and of the invariant solutions found. (a)} Space-averaged temporal spectra of the temperature fluctuations $\langle \vert\hat{\theta}\vert^2\rangle_{x,z}$ as a function of frequency $f$ and $\Ray$. Superimposed solid lines with circle markers track the fundamental frequencies $f_0$ of the converged periodic orbits: PO1 (red) and PO2 (green), while dashed lines with triangles correspond to the Floquet frequencies $f_\mu$. Gray diamond markers indicate specific linear combinations of fundamental and Floquet frequencies, given by $nf_{0}+mf_{\mu}$, with $n, m \in \mathbb{Z}$. 
    \REVA{(b) Existence, stability, and bifurcations of the invariant solutions as a function of $\Ray$. Each row corresponds to one solution, with solid (dotted) lines denoting the ranges where the solution is stable (unstable) and black circles marking the bifurcations.}
    }
    \label{fig:spectrogram}
\end{figure}

We begin by characterizing the behavior of the direct numerical simulations (DNS) of the flow as a function of the Rayleigh number.
\fref{fig:spectrogram}(a) shows the space-averaged temporal spectra of the temperature fluctuation field $\langle \vert \hat{\theta} \vert^2 \rangle_{x,z}$, where $\langle \cdot \rangle_{x,z}$ denotes averaging over the spatial domain, and $\widehat{\cdot}$ is the time Fourier transform. This quantity is shown as a function of frequency $f$ and Rayleigh number $\Ray$. The overlaid markers in the figure and the content of the bottom panel are explained in the next sections.

Several flow regimes can be identified. Except when noted, all regimes found are attractive and unique, meaning every different random initial condition studied eventually converges to the same attractor.
For $\Ray<\num{4.5e5}$, the flow remains in a steady state, signaled by the null region in the frequency domain. 
As $\Ray$ crosses $\num{4.5e5}$, the flow moves away from the steady state and gives rise to periodic motion. The frequency associated to this motion rises steadily as $\Ray$ increases. This is the only observed frequency up until $\Ray = 10^6$, where another transition occurs and secondary frequencies, along with harmonic and combinations thereof, appear in the flow. As the fundamental and new secondary frequencies that emerge are 
incommensurate, the dynamics of the system become quasiperiodic, described by a 2-torus, as shown in Section \ref{sec:characterization}, similar to what is observed in three-dimensions~\cite{puigjaner_steady_2011}. 
This behavior persists in the range $\Ray\in(10^6,\,3.5\times 10^6)$, at the end of which a phase-locking phenomenon takes place, by which the ratio of the two most dominant frequencies becomes rational, with relation 3:1, and the system recovers periodic motion. At $\Ray=4.5\times 10^6$ the flow becomes unstable again and a new quasiperiodic state emerges. 
Afterwards, periodic motion is hard to observe but the overall pattern of the flow jumping from one quasiperiodic state to another keeps repeating, with further transitions occurring at $5.8 \times 10^6$ and $9 \times 10^6$. These changes can be appreciated by the appearance and disappearance of the different branches associated with the various frequencies present. While previous results~\cite{zienicke,paul_bifurcation_2012} had already shown the existence of quasiperiodic states, none had identified the distinct substates reported here. Also, contrary to what is reported for the free-slip case (and at slightly higher Pr)~\cite{paul_bifurcation_2012}, the system never recovers a steady state.
Finally, at around $\Ray=1.4 \times 10^7$ the spectrum becomes more broadband, and as we will see later, this marks the transition to chaos.

As stated above, for each $\Ray$ we initialized the flow from different random initial conditions, which all converged to the same statistically stationary state at each given $\Ray$, except for a narrow band around $\Ray=10^7$. 
\REVB{In this interval, observed between $\Ray=\num{0.9e7}$ and $\num{1.05e7}$, different initial conditions lead to one of two distinct states. Both share the same large-roll structure and dominant frequency, but present different secondary spectral peaks and differ in the amplitude of their enstrophy fluctuations by roughly $40\%$. These persistent dissimilarities
}
suggest the coexistence of distinct but similar attracting solutions consistent with previous observations in related configurations~\cite{paul_bifurcation_2012}.
A detailed characterization of this phenomenon is beyond the scope of the present work.

\subsection{Characterization of invariant solutions}
\label{sec:characterization}

The spectral analysis presented in the previous section suggests the existence of several branches of invariant solutions. We begin by characterizing the spatiotemporal structure and stability of the converged solutions. In the next section we compare their behavior with the observed dynamics in the flow and point out the regions where additional invariant solutions may exist but the Newton-Krylov method failed to find.

\begin{figure}
    \centering    \includegraphics[width=0.95\linewidth]{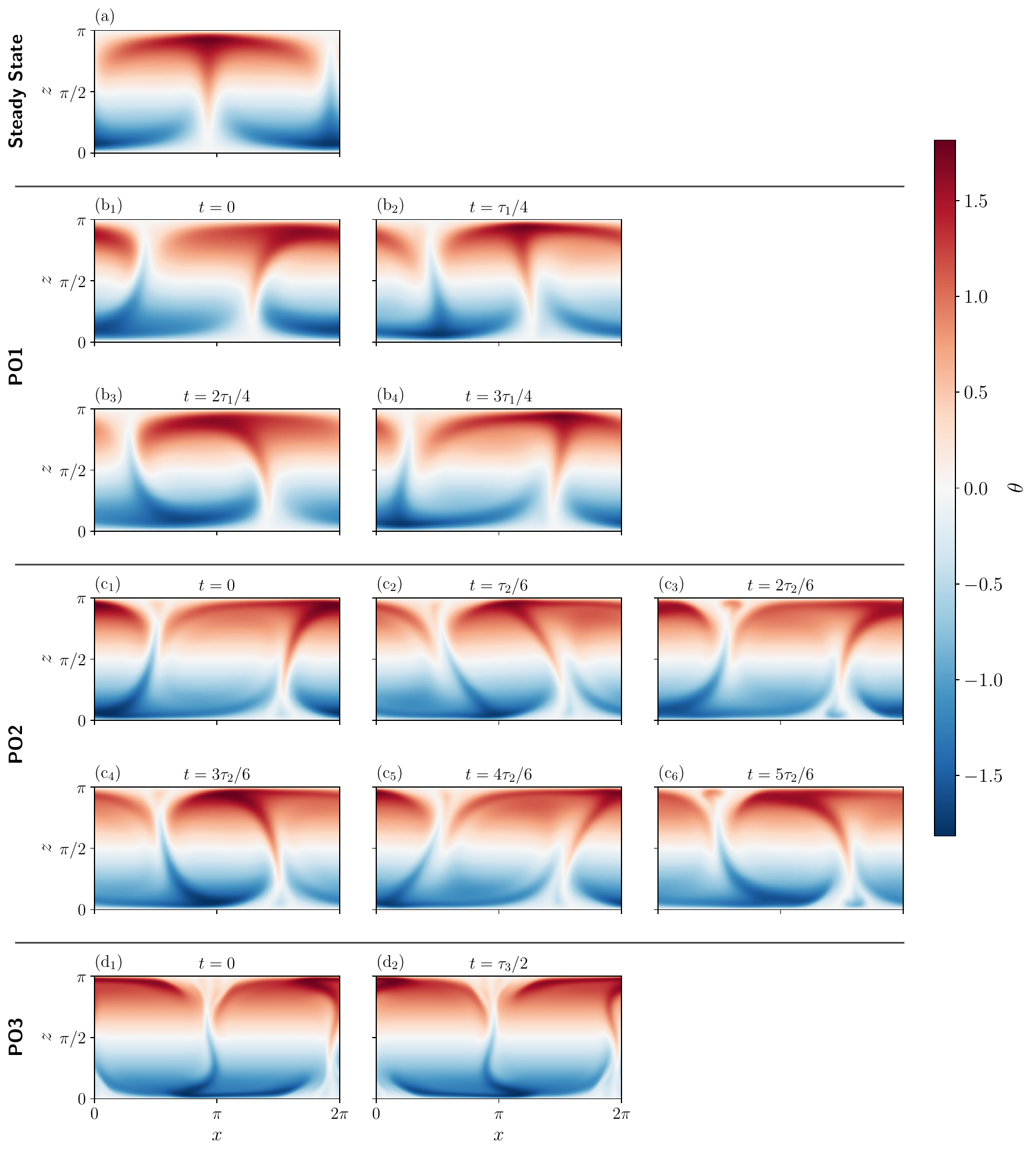}
    \caption{Temperature visualizations of converged invariant solutions.
    \REVA{(a) Steady state ST at $\Ray=\num{4e5}$, consisting of a convective roll with symmetric rising warm and sinking cold plumes.
    $\mathrm{(b_1-b_4)}$ Four equispaced snapshots spanning one period of PO1 at $\Ray= \num{8e5}$ ($\tau_1=11.98$) showing the back and forth swaying motion of the plumes. 
    (c) Six equispaced snapshots spanning one period of PO2 at $\Ray=4\times 10^6$ ($\tau_2=31.17$).
    (d) Two snapshots spanning one period of PO3 at $\Ray=1.5\times 10^7$ ($\tau_3=1.87$).
    }
    }
    \label{fig:viz}
\end{figure}

We found four invariant solutions: one steady state, ST, and three periodic states, PO1, PO2 and PO3. 
\fref{fig:viz} shows their corresponding visualization. 
The steady state (ST) consists of a \REVA{convective roll pair} characterized by two primary counter-rotating vortices. The upwelling of warm fluid (positive $\theta$) and downwelling of cold fluid (negative $\theta$) are perfectly symmetric, as will be discussed in Section~\ref{sec:chaos}. We were able to continue this state throughout the whole range of $\Ray$ under consideration. Its spatial structure remains largely unchanged across this range, but as $\Ray$ increases, the plumes intensify and narrow, leading to steeper gradients in the thermal boundary layers. Similar findings were reported on the structure of the steady state in an elongated box~\cite{kooloth_coherent_2021} .

\begin{figure}
    \centering
    \begin{subfigure}{0.49\linewidth}
        \centering
        \includegraphics[width=\linewidth]{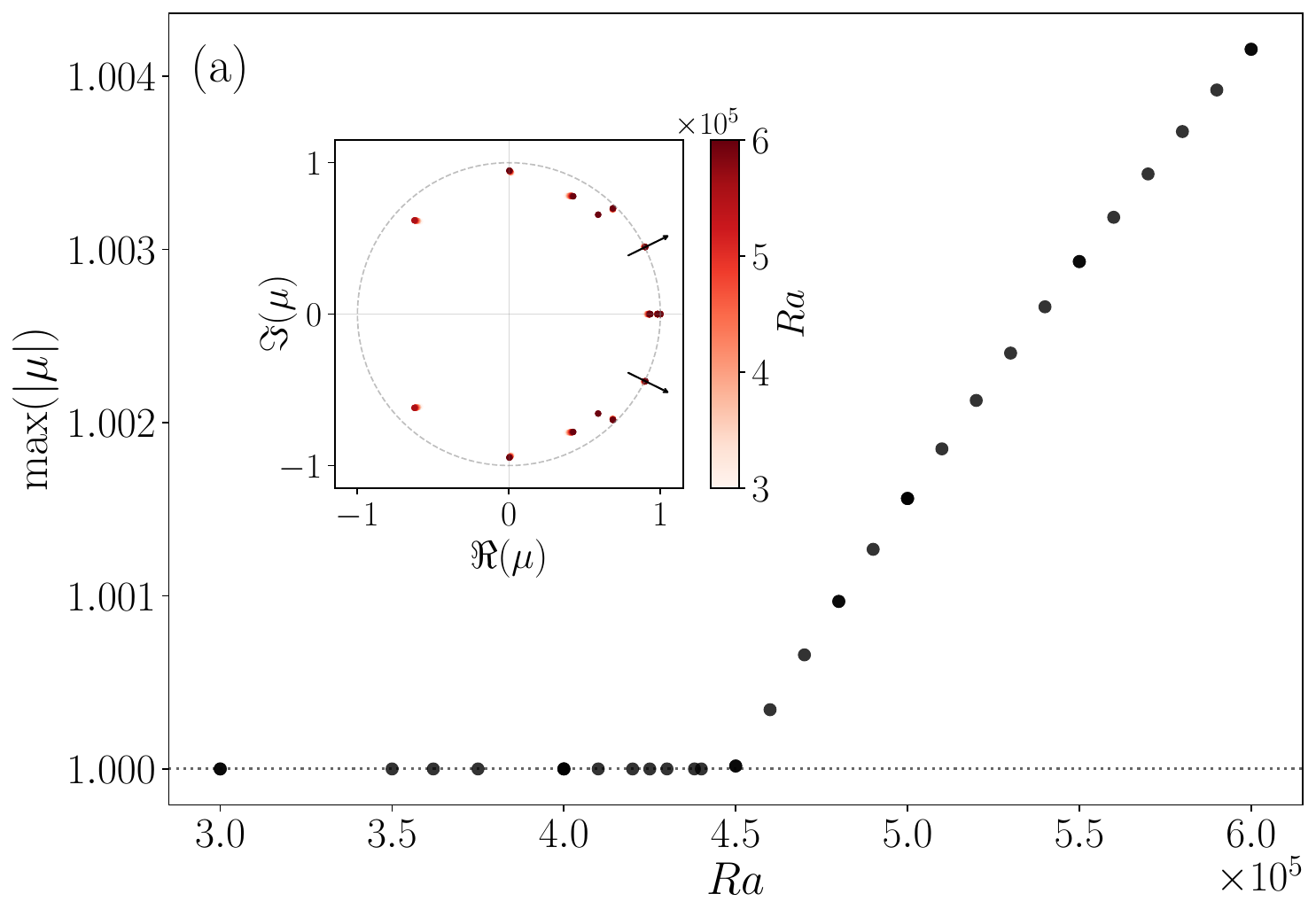}
        \phantomcaption
        \label{fig:floquet_steady}
    \end{subfigure}
    \begin{subfigure}{0.49\linewidth}
        \centering
        \includegraphics[width=\linewidth]{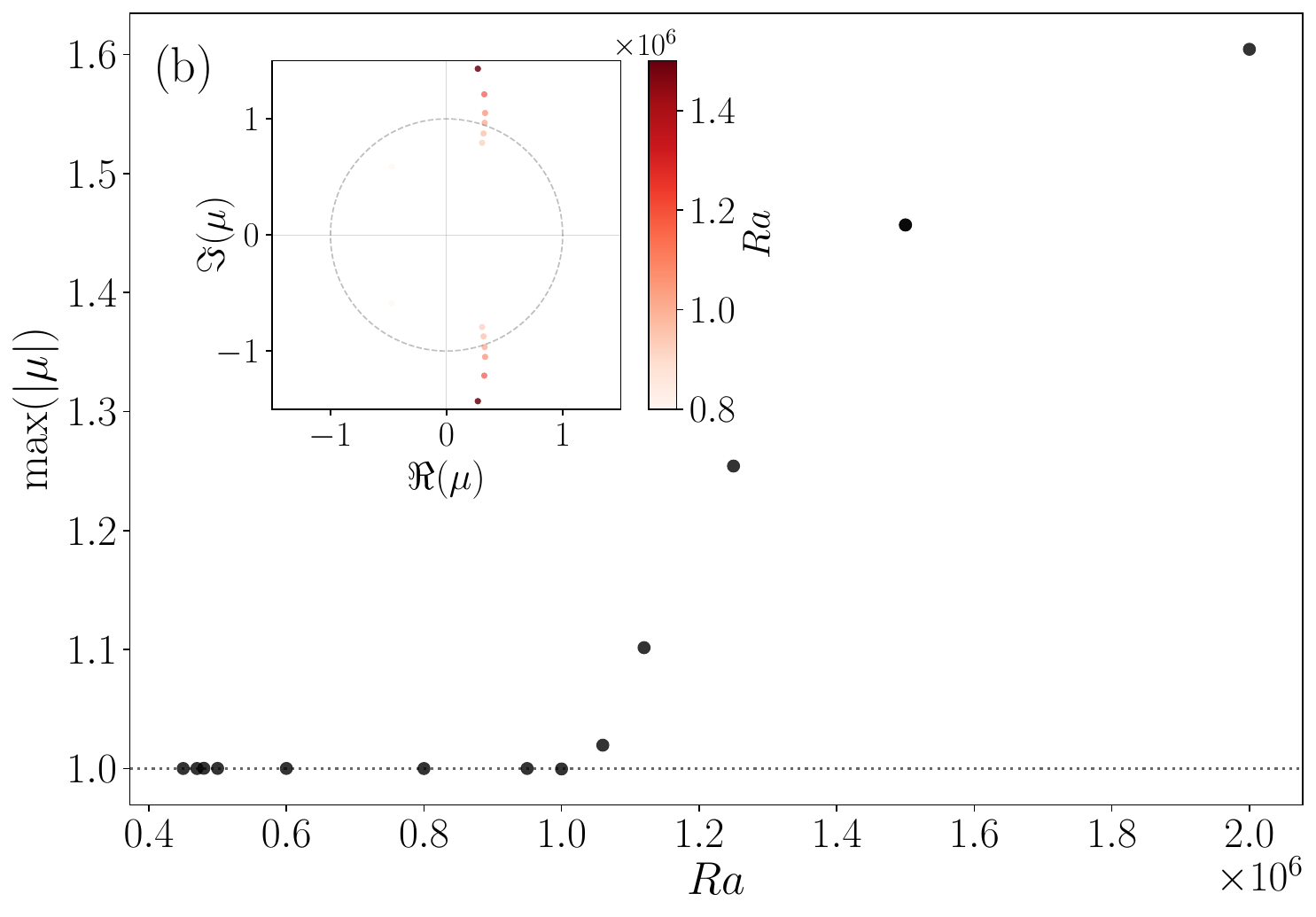}
        \phantomcaption
        \label{fig:floquet_PO1}
    \end{subfigure}
    \begin{subfigure}{0.49\linewidth}
        \centering
        \includegraphics[width=\linewidth]{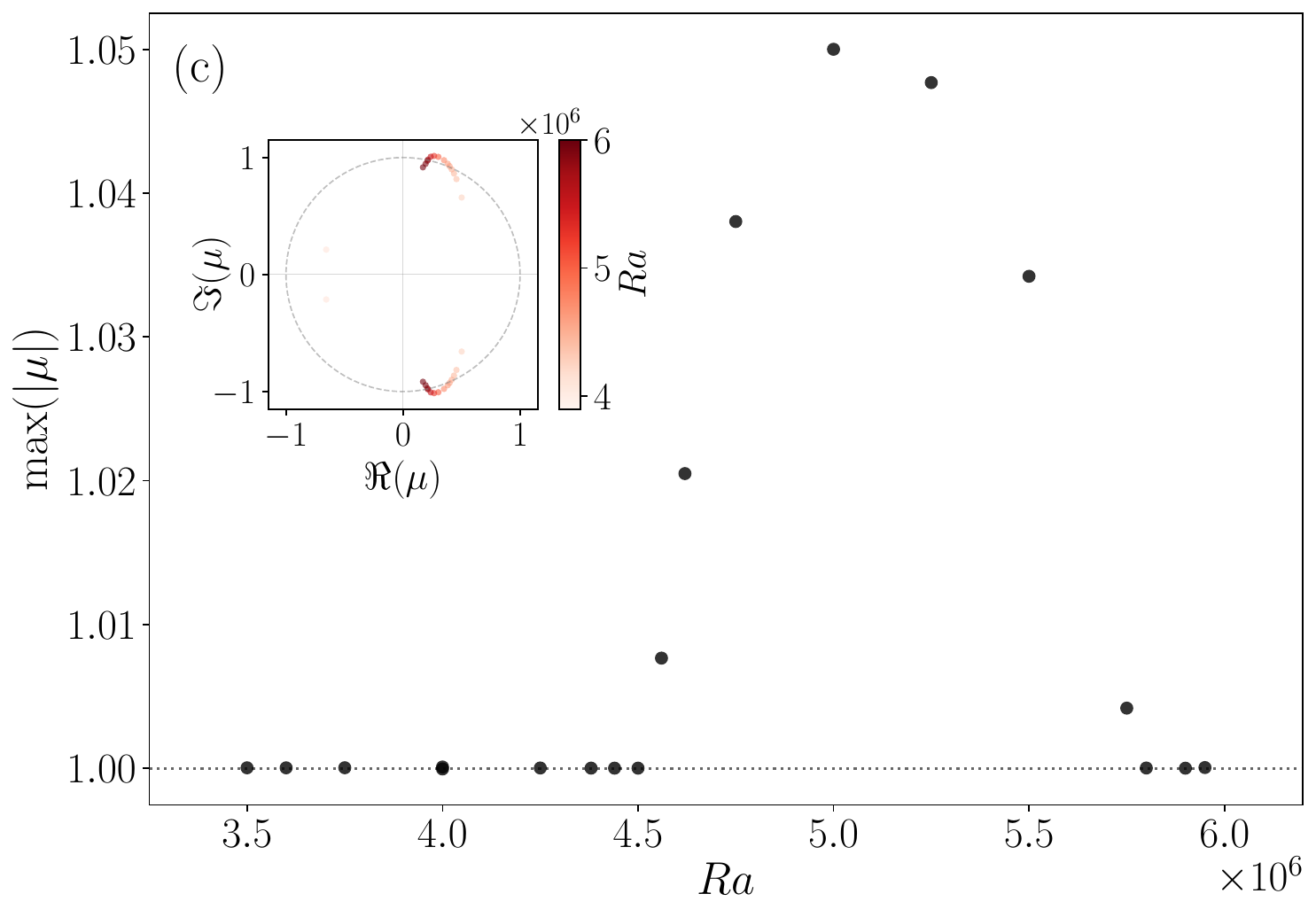}
        \phantomcaption
        \label{fig:floquet_PO2}
    \end{subfigure}
    \caption{Stability of invariant solutions as a function of $\Ray$. (a) ST, (b) PO1, (c) PO2. Periodic orbit figures display magnitude of the leading Floquet multiplier, $\max (|\mu|)$, with the insets showing the corresponding complex Floquet spectra (only showing 2 multipliers of largest magnitude) in the complex plane, with color indicating $\Ray$. For the steady state case eigenvalues of the flow-map Jacobian are shown, displaying the top 15 values in the complex plane. }
    \label{fig:floquet}
\end{figure}

\begin{figure}
    \centering
    \includegraphics[width=0.95\linewidth]{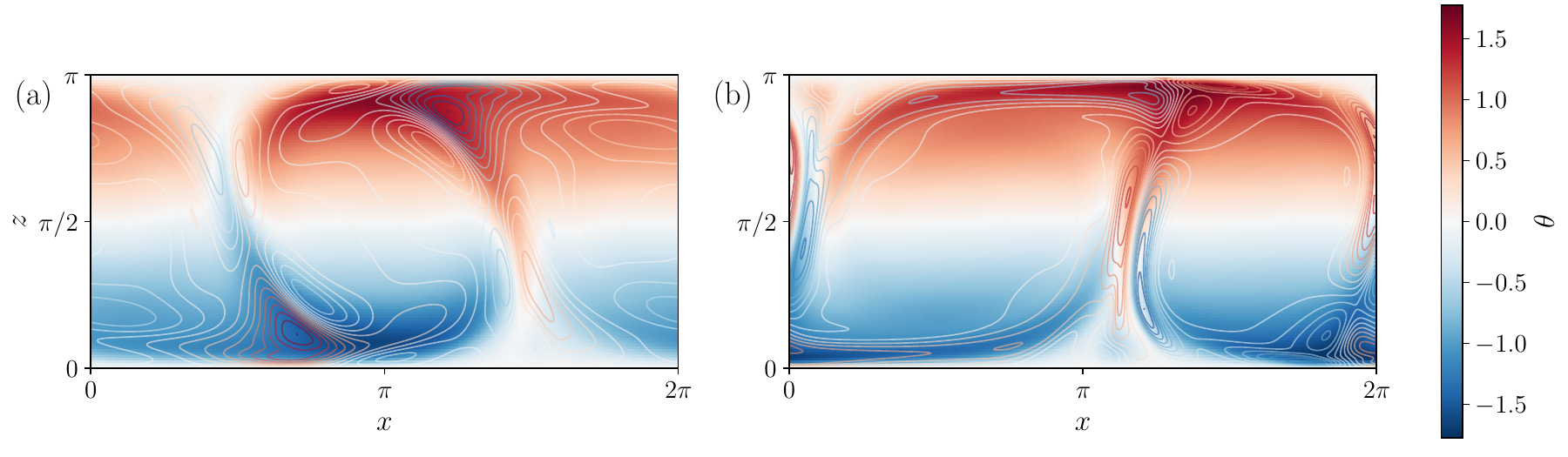}
    \caption{Temperature fluctuation field at $\Ray = 1.06\times 10^{6}$ (a) and $\Ray = 4.5\times 10^{6}$ (b) overlaid with contours of the real part of the most destabilizing Floquet eigenvector. The eigenmode highlights the spatial structure of the leading instability of the periodic orbit.}
    \label{fig:floq_fields}
\end{figure}

The first periodic orbit (PO1), depicted in the second panel of \fref{fig:viz}, consists of a swaying, back and forth motion of the warm and cold plumes, with a period of $\tau_1 = 11.98$  at $\Ray=8\times 10^5$. It is worth noting that the dynamics unfold mainly in the top (bottom) of the warm (cold) plume, while the other end of the plume exhibits only weak motion with small temperature fluctuation. In addition, no mean horizontal displacement is produced over one period ($s=0$). 
This state could be converged from $\Ray = \num{4.5e5}$ (destabilization of ST), up to $\Ray=\num{2e7}$, at which point the continuation failed to produce a converged state. 


The second periodic orbit (PO2) is shown in the third panel of \fref{fig:viz}. This orbit has a period of $\tau_2=31.17$ for $\Ray=4\times 10^6$, roughly 3 times that of PO1, and has no horizontal shift. 
In contrast with the previous solution, the swaying of the plumes cannot be described as a simple back and forth motion, but instead presents a sloshing oscillation, more intricate than the spatiotemporal dynamics of PO1. Additionally, the steady ends of the plumes acquire a non-zero temperature, as can be seen for $t=2\tau_2/6$ and $t=5\tau_2/6$. This state could only be continued in the range $3.5 \times 10^6 \leq \Ray \leq 6 \times 10^6$.

A third periodic orbit (PO3) was also converged, shown in the lower panel of \fref{fig:viz}, in the range $\Ray \in (\num{8e6},\;\num{1.8e7})$ with a period $\tau_3 \approx 2$, an order of magnitude shorter than PO1 and PO2, with no horizontal shift as well. However, this solution is linearly unstable throughout its converged $\Ray$ range, and its short-period dynamics can be understood as a periodic fluctuation about the steady state.
For these reasons we set aside a detailed analysis for PO3, and it is only considered for the Sections \ref{sec:comparison} and \ref{sec:nusselt}.

The linear stability analysis of ST, PO1, and PO2 solutions is summarized in \fref{fig:floquet}, which reports the absolute value of the leading Floquet multiplier max$(|\mu|)$ as a function of $\Ray$. 
In addition, the insets show the leading multipliers in the complex plane. Stability is lost when a multiplier crosses the unit circle.
For ST, shown in \fref{fig:floquet_steady}, the neutral eigenvalue $1.0$ remains the largest up to $\Ray=4.5\times 10^5$, where a complex-conjugate pair becomes critical and exits the unit circle.
This identifies a Hopf bifurcation. 
For PO1 (\fref{fig:floquet_PO1}) the nontrivial multipliers remain inside the unit circle up to $\Ray=10^6$, where a complex pair crosses outward. This is the signature of a Neimark-Sacker bifurcation.
The Floquet spectrum of PO2, shown in \fref{fig:floquet_PO2}, exhibits two changes of stability. A complex pair first leaves the unit circle at $\Ray= 4.5\times 10^6$, marking another Neimark-Sacker bifurcation, and later re-enters for $\Ray\in(5.7\times 10^6,\,6\times 10^6)$, delimiting the interval in which PO2 is linearly unstable. As stated above, this solution could not be continued any further. These results are summarized in the bottom panel of \fref{fig:spectrogram}, where solid lines represent the stable regions of each solution and dotted lines represent the unstable ones. 


Finally, in \fref{fig:floq_fields}(a) we show the isocontours of the leading unstable Floquet modes overlaid on the converged temperature field for a Rayleigh number just beyond the Neimark-Sacker transition ($\Ray = \num{1.06e6}$). 
This provides insight regarding the instability mechanism driving the bifurcation, as the extremes of the Floquet vectors can be identified as the most sensitive region of the flow. 
It is clear that the spatial structures of these extreme values are centered around the hot and cold plumes, as was also seen for the imaginary part (not shown). 
It suggests that the most unstable perturbation arises from an antisymmetric disturbance from the convective roll.
Similarly, for the transition at $\Ray = \num{4.5e6}$, \fref{fig:floq_fields}(b) showcases comparable behavior, with the instability concentrated primarily at the center of the plumes.
\REVA{Furthermore, as we show in Section~\ref{sec:comparison}, the new frequencies appearing in the DNS after each bifurcation match the Floquet frequencies of the corresponding orbit, so the quasiperiodic flow can be depicted as the orbit modulated by these spatial modes.}

\subsection{Comparison between invariant solutions and observed dynamics}
\label{sec:comparison}

Again, we begin with the lower range of Rayleigh numbers studied. 
As mentioned above, the flow transitions from steady to periodic motion at $\Ray=\num{4.5e5}$. This transition coincides with the Hopf bifurcation experienced by solution ST. The first Floquet frequency of ST is shown in \fref{fig:spectrogram} with open blue triangle markers, while the fundamental frequency of PO1 is shown with solid red circular markers.

\begin{figure}
    \centering
    \begin{subfigure}{\linewidth}
        \centering
        \includegraphics[width=0.82\linewidth]{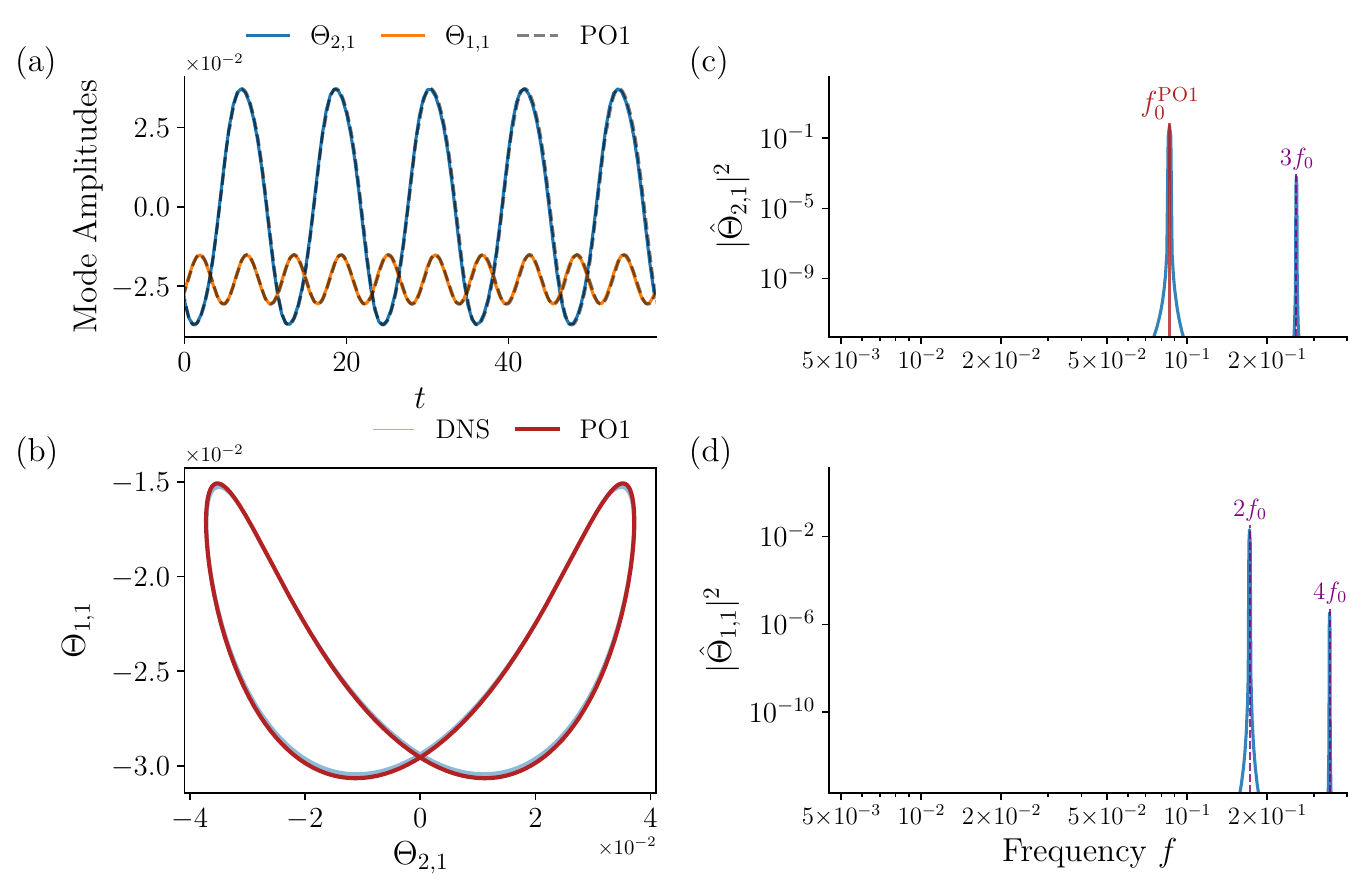}
        \phantomcaption
        \label{fig:viz_1e6}
    \end{subfigure}
    \begin{subfigure}{\linewidth}
        \centering
        \includegraphics[width=0.82\linewidth]{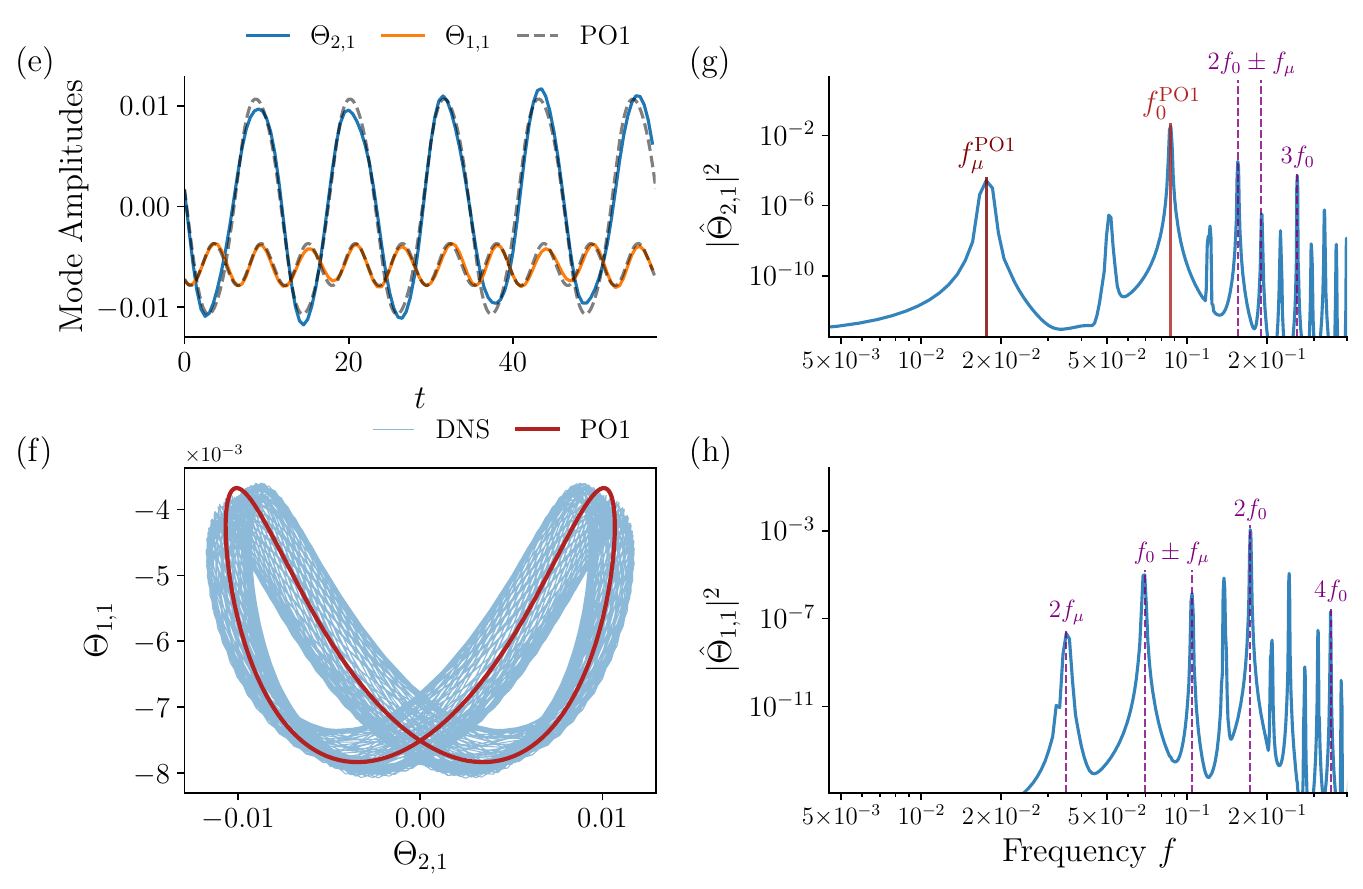}
        \phantomcaption
        \label{fig:viz_1.12e6}
    \end{subfigure}
    \caption{\REVB{Comparison between PO1 and observed full flow dynamics at (a-d) $\Ray = 10^6$, where PO1 is stable, and (e-h) $\Ray = 1.12 \times 10^6$, just after Neimark-Sacker bifurcation. 
    (a) and (e) Time series of the dominant Fourier modes for the DNS (solid lines) and PO1 (dashed). 
    (b) and (f) Projection onto the $({\Theta}_{2,1}, {\Theta}_{1,1})$ plane, comparing DNS (light blue) and PO1 (red).
    (c) and (g) Temporal spectrum of $\vert \widehat{\Theta}_{2,1} \vert^2(f)$, and (d) and (h) of $\vert \widehat{\Theta}_{1,1} \vert^2(f)$, with selected frequencies of PO1 that appear in each spectrum marked.}} 
    \label{fig:first_viz}
\end{figure}

The spatiotemporal correspondence between the converged invariant solutions and the full flow dynamics is demonstrated in \fref{fig:first_viz}(a-c) for $\Ray=10^6$. We analyze the real part of temperature Fourier modes defined as
\begin{equation}
    \Theta_{n,m} = \Re \left[\int_0^{L}\int_0^{h} \theta(x,z)\,\exp\left(2\pi i \frac{nx}{L}\right)\, \sin\left(\pi  \frac{mz}{h}\right)\,dx\,dz\right].
\end{equation}
Our analysis focuses on the dominant large-scale modes, $\Theta_{1,1}$, and $\Theta_{2,1}$.
The time series in \fref{fig:first_viz}(a) of these modes reveal that the periodic orbit (dashed line) closely shadows the DNS trajectory (solid line). 
This is further reflected in the phase-space projection onto the $(\Theta_{1,1}, \Theta_{2,1})$ plane in \fref{fig:first_viz}(b), where the DNS follows a path nearly identical to the invariant orbit.
\REVB{The projection is mirror-symmetric about the $\Theta_{2,1}=0$ axis because of the space-time symmetry of the orbit discussed in Section~\ref{sec:chaos}, under which $\Theta_{2,1}$ changes sign while $\Theta_{1,1}$ is unchanged after half a period.}
The power spectra in Fig.~\fref{fig:first_viz}(c,d) confirm this correspondence, as the DNS frequency peaks match the fundamental frequency and harmonics of PO1. \REVB{The same symmetry restrict $\Theta_{2,1}$ to odd harmonics of $f_0$ and $\Theta_{1,1}$ to even ones, so the spectra of both modes are shown, each with the frequencies of the orbit that appear on it marked.}

Right after $\Ray=10^6$ the PO1 undergoes a Neimark-Sacker bifurcation and the flow becomes quasiperiodic. The red open triangular markers in \fref{fig:spectrogram} denote the Floquet frequency of PO1, while the gray markers show combinations of the fundamental and Fourier frequencies. The frequencies observed in the DNS are well-characterized by those coming from the PO1. This transition is further exemplified in \fref{fig:first_viz}(e-h). The power spectrum \fref{fig:first_viz}(g) signals the emergence of a new relevant frequency, matching the Floquet frequency $f_\mu^{\mathrm{PO1}}$. In \fref{fig:first_viz}(f) a torus can be seen to fill the phase-space, a hallmark of quasiperiodic behavior. Consequently, the DNS dynamics can be interpreted as the invariant orbit modulated by this Floquet frequency, as evidenced by the phase-space projection and the time series of the Floquet coefficients.
As $\Ray$ grows, the dynamics are still reminiscent of a modulated PO1, albeit the fundamental frequency of PO1 begins to drift from the dominant peak of the DNS power spectral density. This frequency decoupling indicates that while PO1 remains an exact solution to the governing equations, it no longer represents the most energetically dominant structure of the flow as the attractor becomes more complex.

\begin{figure}
    \centering
    \begin{subfigure}{\linewidth}
        \centering
        \includegraphics[width=0.82\linewidth]{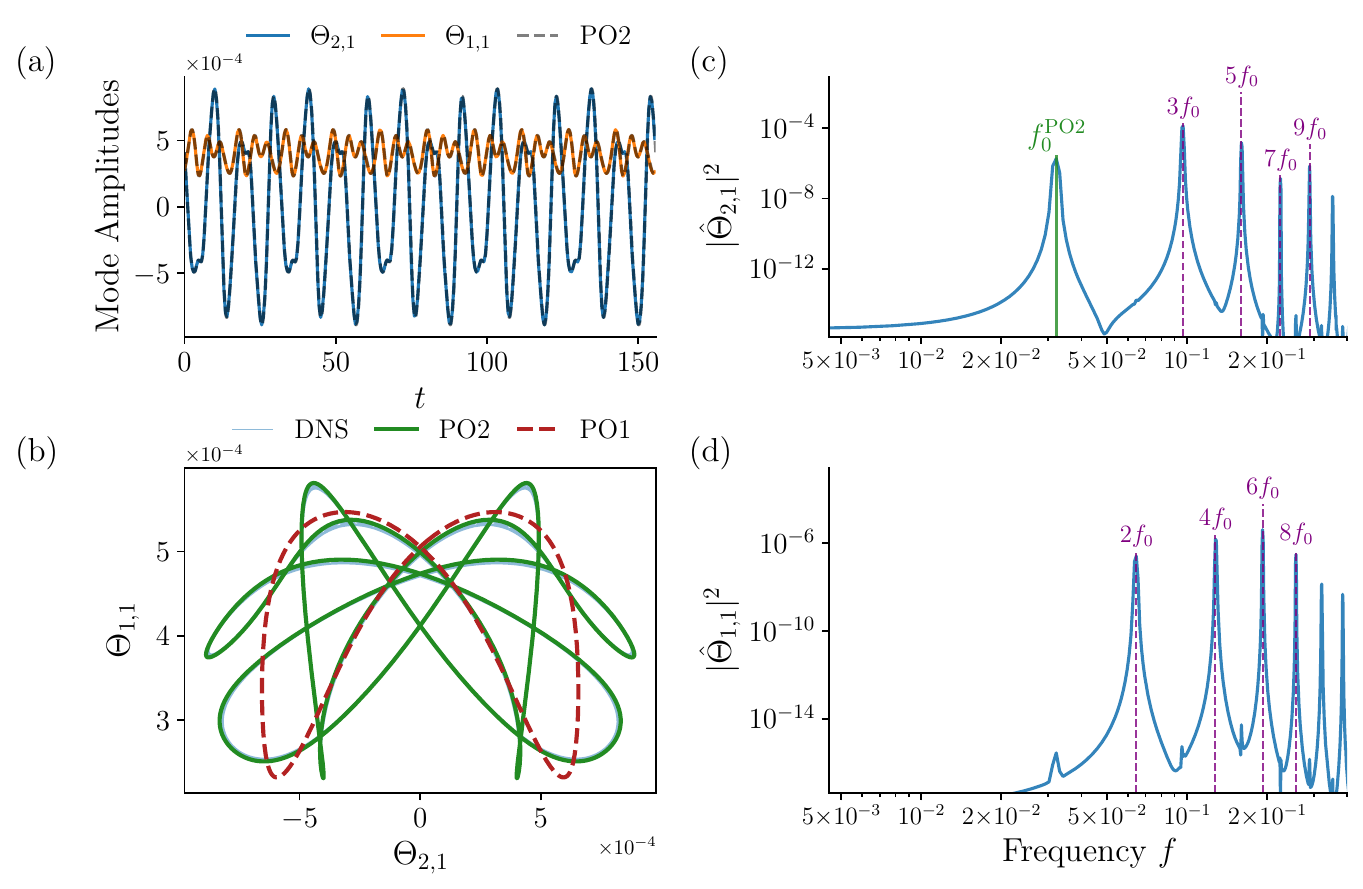}        
        \phantomcaption
        \label{fig:viz_3.75e6}
    \end{subfigure}

    \begin{subfigure}{\linewidth}
        \includegraphics[width=0.82\linewidth]{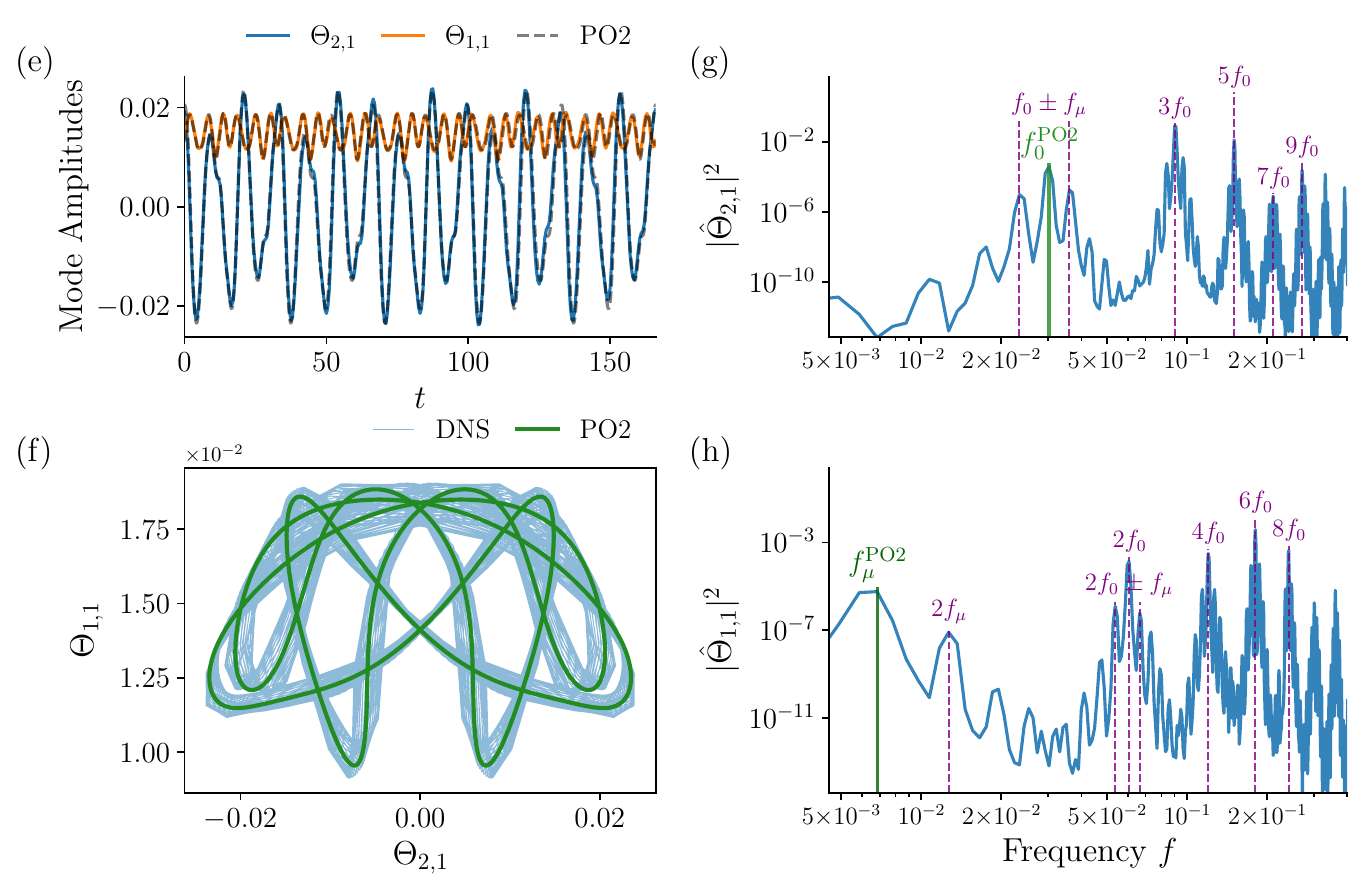}
        \phantomcaption
        \label{fig:viz_5.5e6}
    \end{subfigure}
    \caption{As in \fref{fig:first_viz}\REVA{, now comparing PO2 with the DNS at (a-d) $\Ray = 3.75 \times 10^6$,
    in the phase-locked regime where PO2 is stable,
    and (e-h) $\Ray = 5.5 \times 10^6$
    , where PO2 is unstable and the flow is quasiperiodic.
    Panel (b) also displays PO1 (dashed red line) to highlight how the DNS trajectory follows PO2 instead of PO1.}}
    \label{fig:viz_2po}
\end{figure}

At $\Ray=3.75\times 10^6$, the different frequencies become commensurate and the system enters a phase-locked state that gives rise to PO2. 
In \fref{fig:viz_2po}(a-d) a clear contrast between PO1 and PO2 is evident, the phase-space reveals how the DNS trajectory closely follows PO2, while PO1 presents a simpler stray curve. 
Interestingly, the fundamental frequency of PO2 (marked with green circles in \fref{fig:spectrogram}), is one third of the most energetic frequency in the flow. This shows that while a periodic orbit may fully characterize the dynamics, it may not always do so in the most straightforward fashion. 

After PO2 undergoes another Neimark-Sacker bifurcation, the flow recovers quasiperiodic behavior, as seen in \fref{fig:viz_2po}(e-h). The new frequencies observed match the Floquet frequencies from PO2 or linear combinations of its harmonics.
\REVB{In contrast with PO1, the Floquet frequency of PO2 appears by itself in the spectrum of $\Theta_{1,1}$, \fref{fig:viz_2po}~(h), and only through the sidebands $f_0\pm f_\mu$ in that of $\Theta_{2,1}$, \fref{fig:viz_2po}~(g), since the leading Floquet mode of PO2 is even under the space-time symmetry of the orbit whereas that of PO1 is odd. 
The projection in \fref{fig:viz_2po}~(b) and (f) are both miror-symmetric about the $\Theta_{2,1}=1$ axis as before, consequence of the space-time symmetry of the orbit. 
}

\begin{figure}
    \centering    \includegraphics[width=0.8\linewidth]{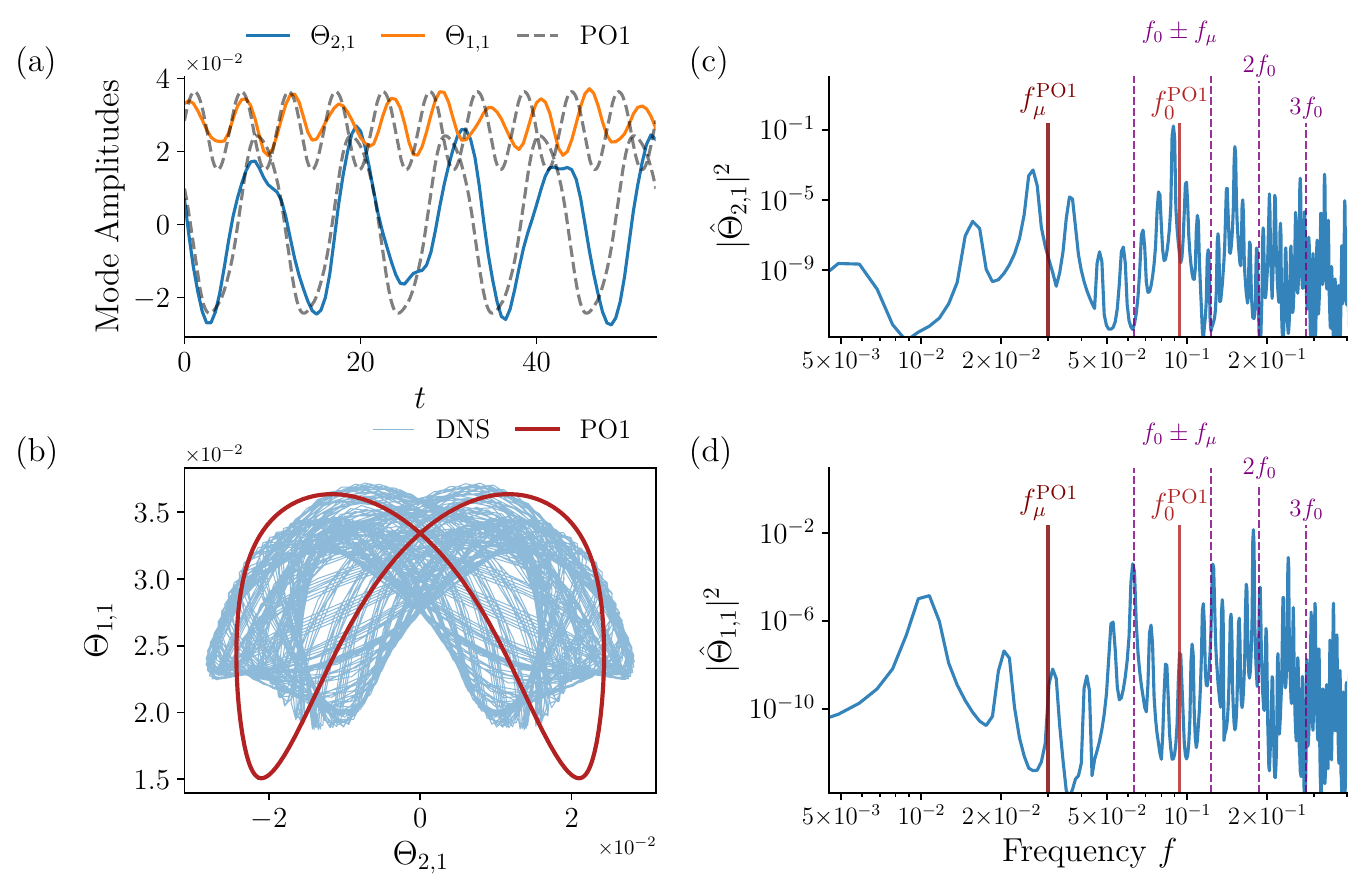}
    \caption{As in \fref{fig:first_viz} with $\Ray = 6.5 \times 10^6$,
    \REVA{comparing the DNS with PO1, the closest converged solution at this $\Ray$.
    (a) Time series of the dominant Fourier modes, (b) projection onto the $(\Theta_{2,1},\Theta_{1,1})$ plane, and (c) and (d) temporal spectra of $\Theta_{2,1}$ and $\Theta_{1,1}$ with the fundamental, Floquet, and harmonics of PO1 marked.}}    
    \label{fig:viz_6.5e6}
\end{figure}

Further increasing the Rayleigh number to $\Ray=6.5\times 10^6$, as shown in \fref{fig:viz_6.5e6}, leads to a regime that still shows clear signs of being close to an unstable orbit. This orbit though is neither PO1, although it is remarkably similar, or PO2, which fails to converge in this range. We tried several initial conditions for the Newton-Krylov method, which involved applying a temporal band-pass filter the fields, but none yielded any results. Our interpretation is that the actual orbit probably has a fundamental frequency around $0.16$, so Newton-Krylov method, which is basically a shooting method, is failing to converge orbits with such long periods.


\begin{figure}
    \centering    \includegraphics[width=0.65\linewidth]{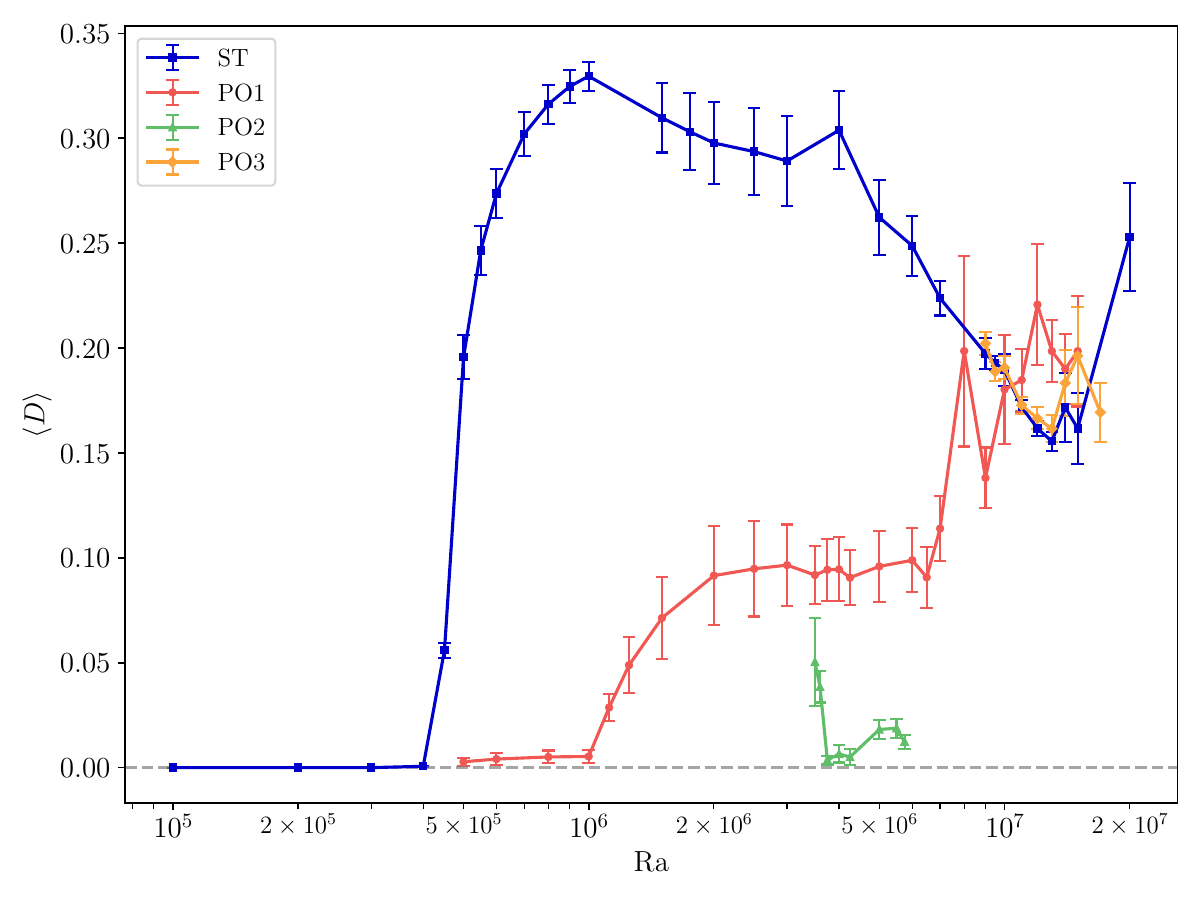}
    \caption{Time-averaged shadowing relative distance $\langle D \rangle$ between the DNS trajectories and each converged invariant solution as a function of the Rayleigh number $\Ray$. \REVB{The error bars corresponds to the standard deviation over time of the instantaneous signal $D(t)$.}} 
    \label{fig:shadowing}
\end{figure}

To quantify how closely each invariant solution shadows the flow dynamics, we define a shadowing distance $D(t)$ between a DNS snapshot at a time $t$ and an invariant solution $\boldsymbol{X}_{\mathrm{IS}}$. This is achieved by minimizing the state-space distance over both the internal phase $t'$ (for periodic orbits) and horizontal shift $s$~\cite{crowley_turbulence_2022}:
\begin{equation*}
    D(t) = \min_{t',\,s}\; \frac{\Vert \mathcal{T}_{s}\,\boldsymbol{X}_{\mathrm{IS}}(t') - \boldsymbol{X}_{\mathrm{DNS}}(t) \Vert}{\Vert \boldsymbol{X}_{\mathrm{DNS}}(t) \Vert}.
    \label{eq:shadowing*}
\end{equation*}
\fref{fig:shadowing} reports the average of this distance over $500$ free-fall time units, calculated after transients have decayed. 
As expected, the distance is near zero during the stable regimes of the invariant solutions, with small deviations resulting from the finite temporal sampling of the fields. 
The highest relative distance is achieved by the unstable ST state, peaking at the Neimark-Sacker bifurcation at $\Ray=10^6$.
At this point, PO1 becomes unstable and the flow starts to deviate from it, though it still remains closer than the steady solution. 
Further on, a sharp increase can be seen between $\Ray=\num{7e6}$ and $\num{8e6}$, near a symmetry breaking event, as will be discussed in \ref{sec:chaos}.
Around this region, the minimum distance to the DNS corresponds to PO2, which persists in close proximity to the flow even within its unstable regime.
On the other hand, PO3 displays a mean distance closely matching that of ST,
further validating the initial viewpoint of the periodic orbit as a perturbation
about the steady state. As stated above, neither ST, PO1 or PO3 are dynamically
relevant after $\Ray = \num{6.5e6}$, serving as another indication that not all
the invariant solutions found in a flow are
meaningful~\cite{cleary_Dynamical_2025,redfern_Dynamically_2024}. It is worth
mentioning that, even though the relative distance appears to decrease with
$\Ray$, the absolute distance increases for all studied solutions.

\subsection{Route to chaos}
\label{sec:chaos}

We now examine the transition to chaos by combining symmetry considerations with a Lyapunov exponent analysis. The discussed spatial transformations, defined by \eqref{eq:syms}, were applied on both the converged invariant solutions as well as the DNS to uncover the symmetries of the flow as $\Ray$ varies, considering the equations were solved with no imposed symmetry contraints.  

\begin{figure}
    \centering    
    \begin{subfigure}{0.49\linewidth}
        \centering        
        \includegraphics[width=\linewidth]{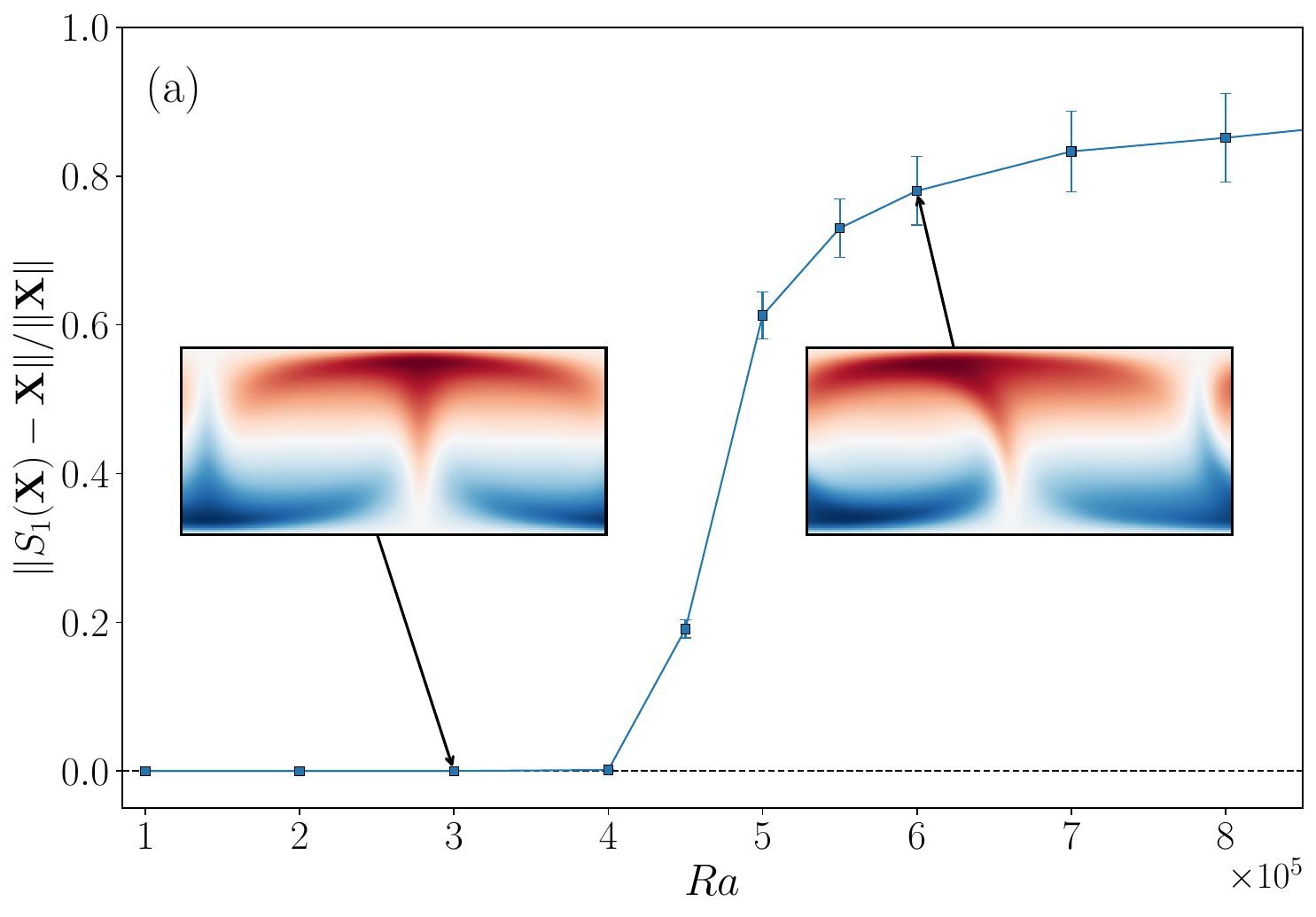}
        \phantomcaption
        \label{fig:symm_S1}
    \end{subfigure}
    \begin{subfigure}{0.49\linewidth}
        \centering        
        \includegraphics[width=\linewidth]{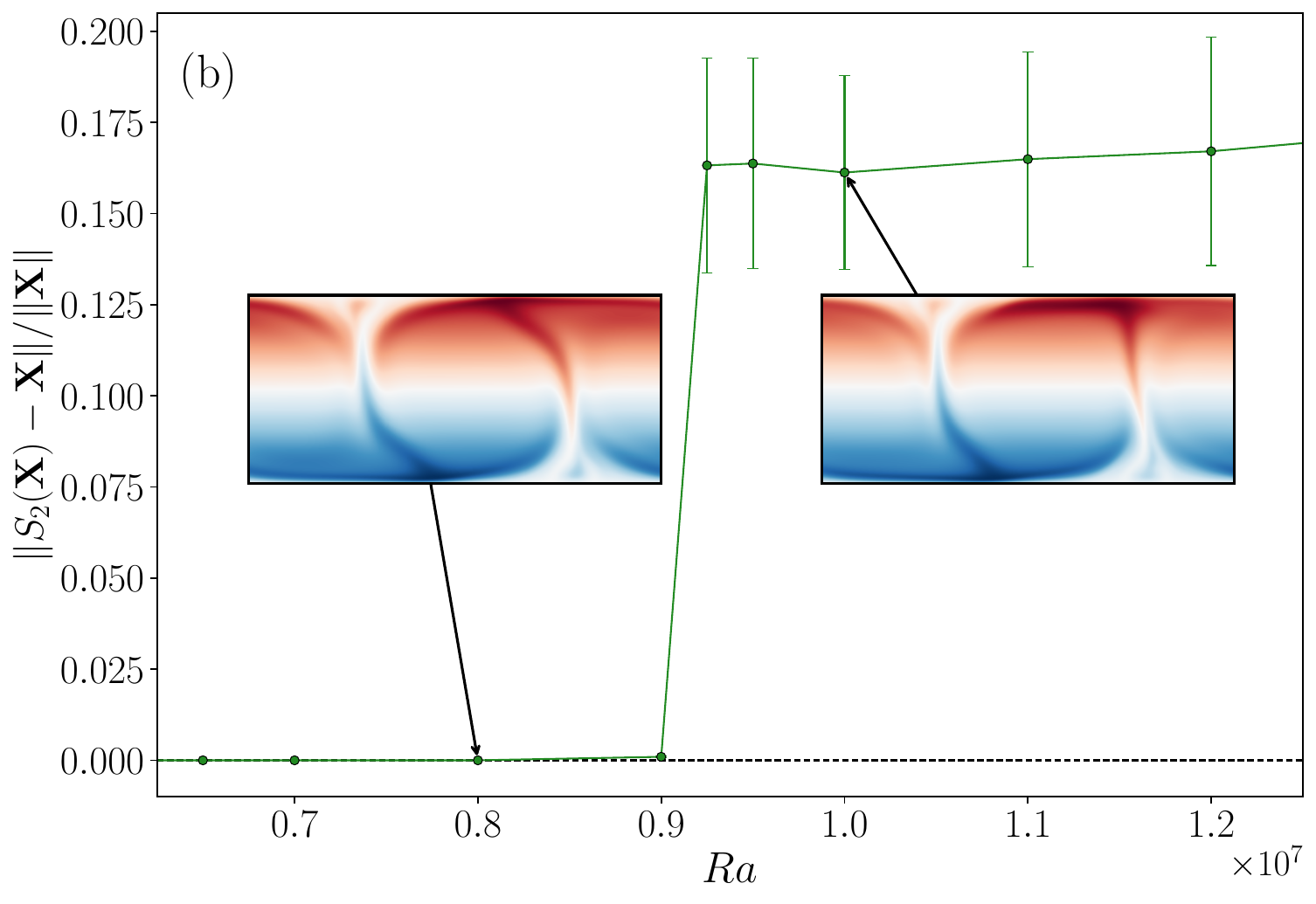}
        \phantomcaption
        \label{fig:symm_S2}
    \end{subfigure}    
    \caption{\REVA{Symmetry breaking of $S_1$ (a) and $S_2$ (b)}. 
    Bifurcation diagram showing the relative error of performing symmetry transformations on DNS fields versus $\Ray$, averaged across time \REVA{after reaching a statistically stationary state. The error bars correspond to the standard deviation of this signal.}
    Dashed lines denote symmetry-preserving solutions. 
    Insets show representative temperature fields below and above the critical $\Ray$, illustrating the transition from a symmetric state to an asymmetric branch.
    }
    \label{fig:symmetry}
\end{figure}

In the stable steady regime, the system preserves all three symmetries, displaying both a horizontal reflection symmetry respect the warm plume centerline ($S_1$) and a vertical reflection combined with a horizontal shift ($S_2$). 
As the system undergoes the Hopf bifurcation, there is a symmetry breaking of $S_1$, as illustrated in \fref{fig:symmetry}(a). The periodic tilting of the plumes disrupts the reflection symmetry, while preserving $S_2$, signaling that the behavior of the warm plume is mirrored by the cold plume and vice-versa.
This symmetry breaking mechanism is further supported by the study of the Floquet eigenvectors of ST, which exhibit an anti-reflection symmetry with respect to the plume centerline, opposing $S_1$, as well as the opposite transformation of $S_3$ (same spatial action reversing sign of fields).

Moreover, analysis of the the converged PO1 reveals that $S_1$ is actually preserved if we compare a snapshot of the fields with the evolution by half a period ($\tau_1/2$), which was also found by~\cite{zienicke} as a space-time symmetry. 
The Floquet eigenvectors of PO1 exhibit the corresponding anti-$S_1$ symmetry under the same half period translation.
PO2 presents the same behavior: it exhibits $S_2$ symmetry and $S_1$ symmetry under a half-period time shift. 
\REVB{Combining the two, both orbits are also invariant under $S_3$ followed by a half-period time shift, which for the modes used in Section~\ref{sec:comparison} implies a multiplication of $\Theta_{n,m}$ by $(-1)^{n+m}$.
}
At higher $\Ray$ numbers, the breaking of the $S_2$ symmetry occurred at $\Ray=9\times 10^6$ (\fref{fig:symmetry}(b)), signaling that it was not prompted by the Neimark-Sacker bifurcation nor any of the other previously discussed transitions. 

\begin{figure}
    \centering
    \includegraphics[width=0.7\linewidth]{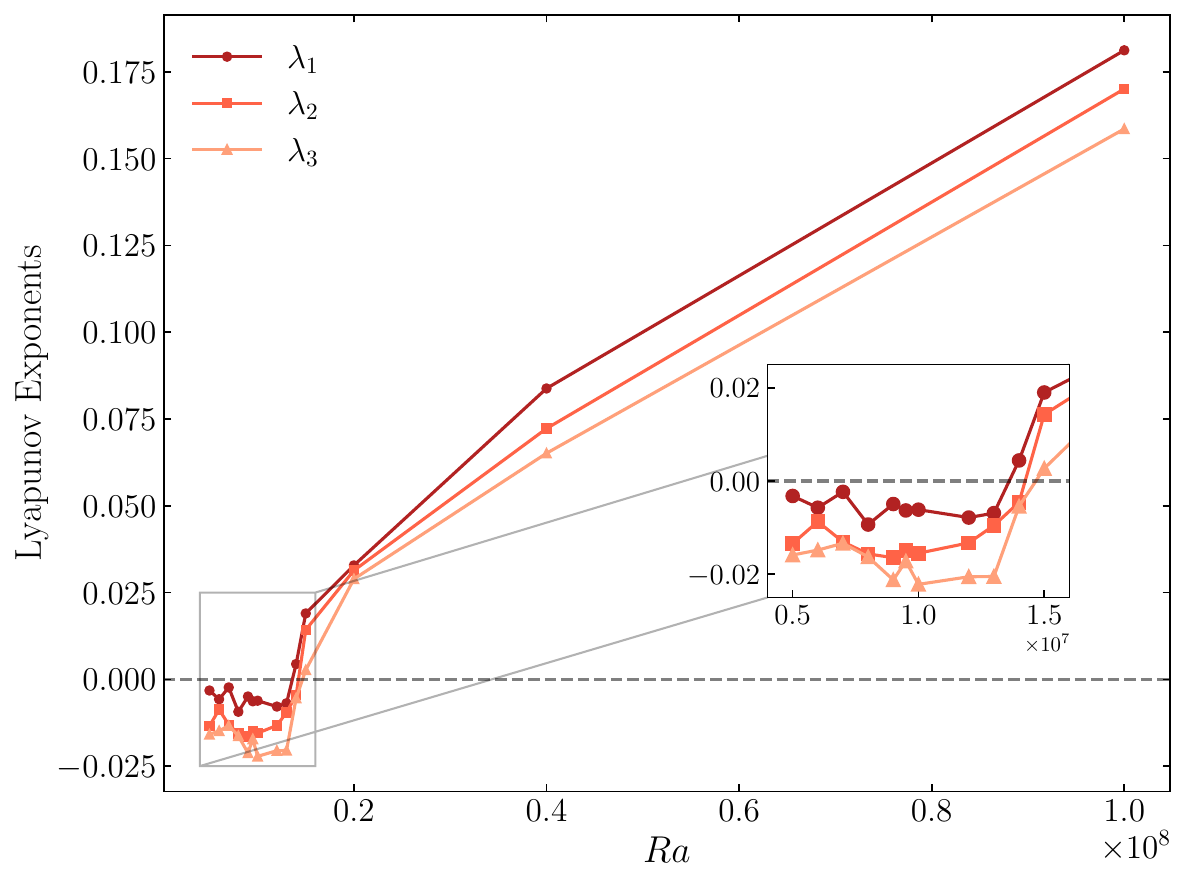}
    \caption{
    Three largest finite-time Lyapunov exponents $\lambda_i$ computed from direct numerical simulations as a function of $\Ray$. 
    The dashed line denotes neutral stability at $\lambda=0$. 
    At low $\Ray$, all exponents are negative, indicating stable dynamics. 
    The leading exponent $\lambda_1$ crosses zero at $\Ray=1.4\times 10^7$, marking the onset of chaos. 
    }
    \label{fig:lyap}
\end{figure}

To quantify the onset of chaos, we computed the finite-time Lyapunov exponents framework by means of the Benettin algorithm~\cite{benettin_lyapunov_1980}.
We calculated the $K=10$ leading exponents with a re-orthonormalization interval of $\Delta t=8$ free-fall times. 
A range of integration times was tested, and the final reported values correspond to the average over the interval where convergence was observed.
The first three leading exponents are shown in \fref{fig:lyap}. 
For $\Ray<10^7$ all exponents remain negative, indicating that the dynamics, although increasingly complex and symmetry-broken, remain non-chaotic. 
The maximal Lyapunov exponent becomes positive at $\Ray=1.4\times10^7$, marking the onset of chaos. This separation between the symmetry-breaking event and the appearance of a positive Lyapunov exponent shows that the route to chaos is not directly triggered by symmetry loss, but rather by the gradual amplification of instabilities within an already asymmetric state.

\begin{figure}
        \centering               \includegraphics[width=0.55\linewidth]{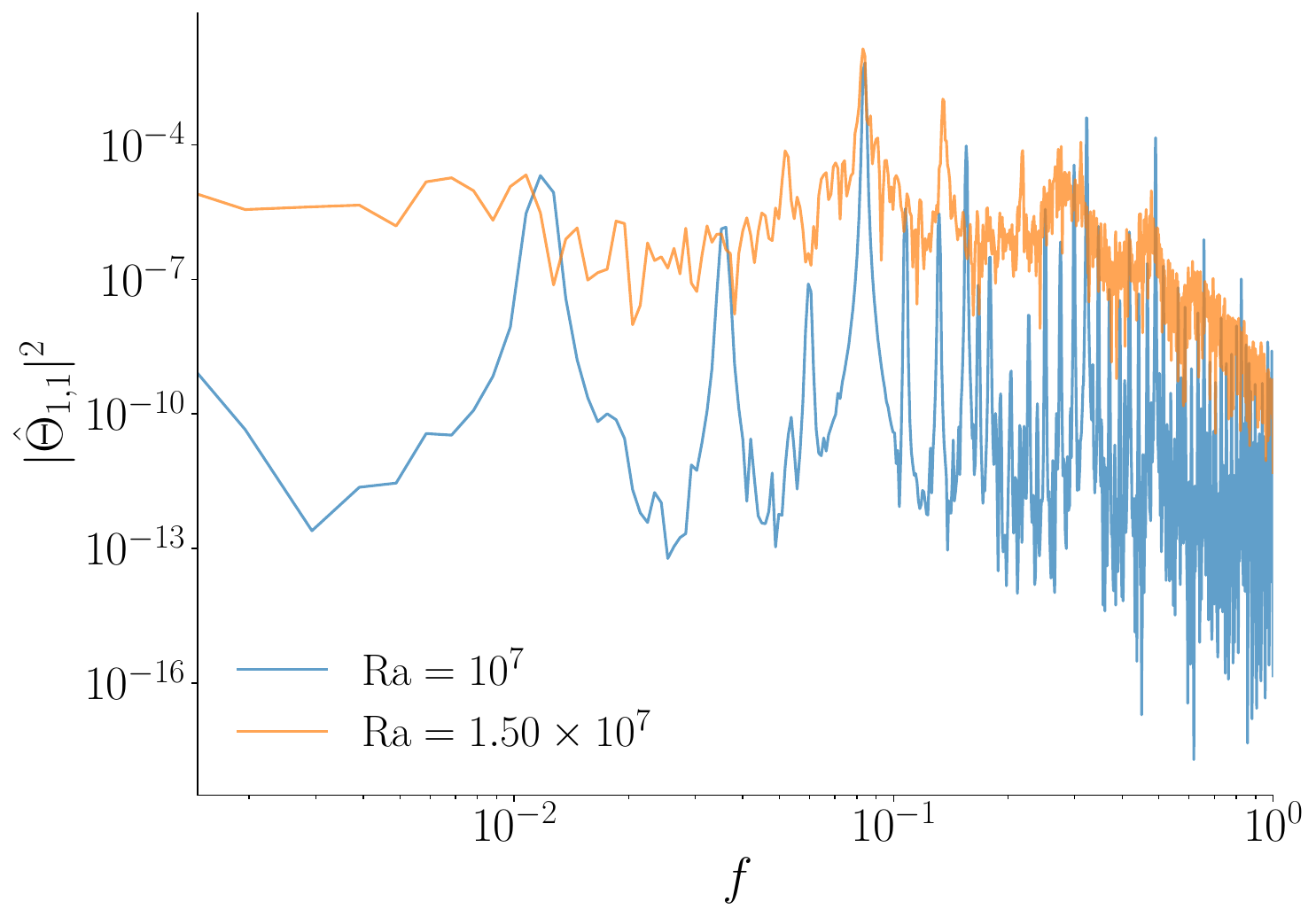}
    \caption{
    Power spectral density of the temperature signal for Rayleigh numbers before and after the chaotic transition.}
    \label{fig:psd}
\end{figure}

The spectral character of this transition is illustrated in~\fref{fig:psd}, which compares the power spectral density of temperature signal at $\Ray=10^7$ and $\Ray=\num{1.5e7}$. Below the chaotic threshold, the spectrum is dominated by a set of discrete peaks at the fundamental frequency and its integer combinations with a secondary frequency, consistent with the quasiperiodic dynamics described in Section~\ref{sec:regimes}. Above the onset of chaos, this discrete structure gives way to a broadband spectrum. Dominant peaks are still present, but on top of an elevated noise floor across the frequency range.

\subsection{Heat transport \REVB{and energy} analysis}
\label{sec:nusselt}

\begin{figure}
    \centering    
    \includegraphics[width=\linewidth]{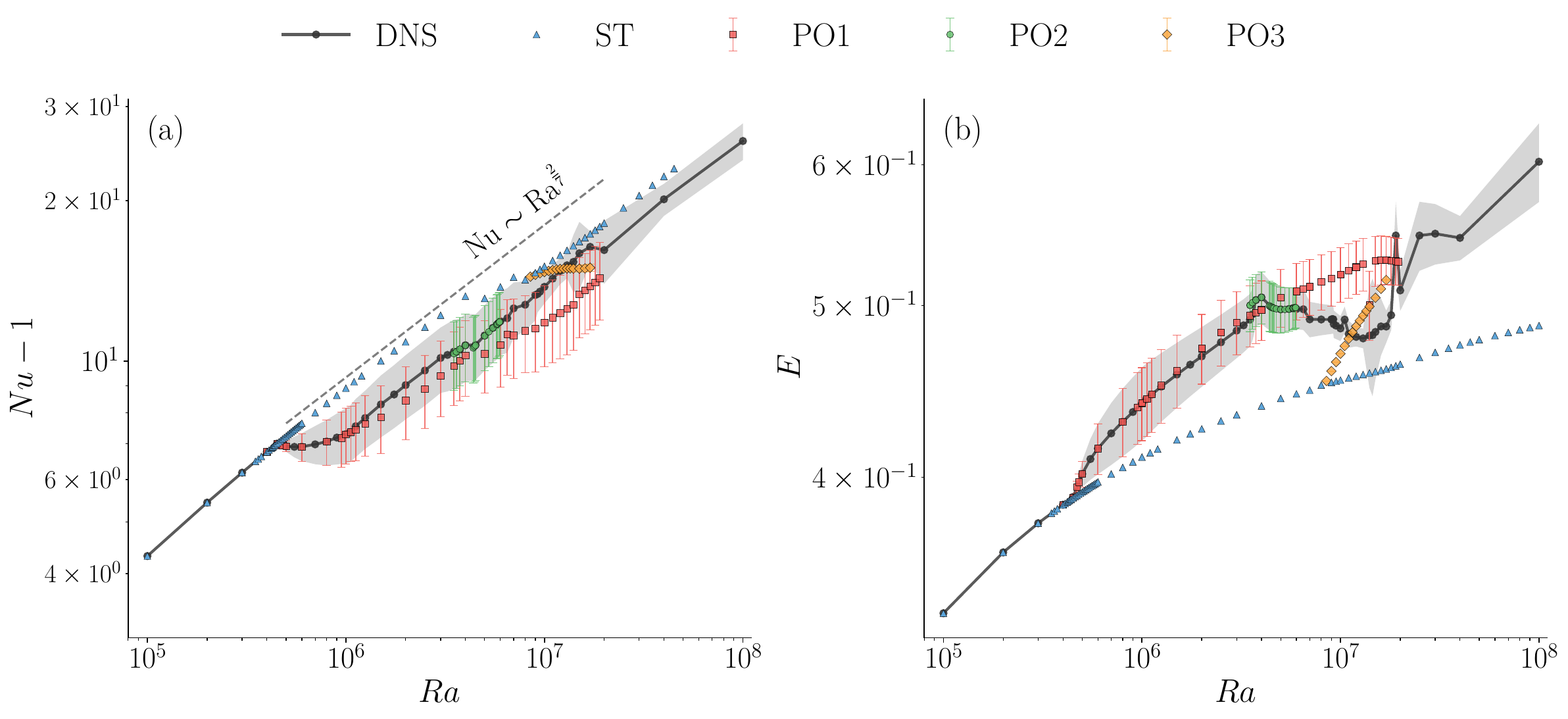}
    \caption{Global heat transport \REVB{and energy}. 
    \REVA{\REVB{(a)} $\Nuss-1$ (ratio between convective and conductive heat transport) as a function of $\Ray$ for the fully developed flow (DNS, dark solid line) and the computed invariant solutions ST (blue triangles), PO1 (red squares), PO2 (green circles), and PO3 (orange diamonds). The gray shaded region around the DNS curve indicates the amplitude of the temporal fluctuations of the $\Nuss$ through its standard deviation over the last 500 free-fall time units, and the error bars on the periodic-orbit markers span the standard deviation across one period of each orbit. 
    The dashed line displays the $\Nuss \sim \Ray^{2/7}$ scaling for reference. } \REVB{(b) Energy as a function of $\Ray$ for the same states. The shaded region and error bars are defined as in (a).}
    }
    \label{fig:nusselt-energy}
\end{figure}

In \fref{fig:nusselt-energy}(a) we show $\Nuss-1$ (ratio between convective and conductive heat flux) as a function of $\Ray$ for the DNS and the four invariant solutions (ST, PO1, PO2, PO3)\REVB{, while \fref{fig:nusselt-energy}(b) shows the kinetic energy of the velocity fields, $E=\tfrac{1}{2} \langle \boldsymbol{v} ^2\rangle_{x,z}$  for the same states}. 
Similar to previous findings~\cite{johnston_Comparison_2009}, there is a sharp transition \REVB{in the heat transport} when the flow becomes periodic, followed by a smooth transition with the onset of quasiperiodicity. 

Considering values of $\Ray>10^6$ after the transition, the observed scaling for the DNS yields a fit of $\Nuss \approx 0.16\,\Ray^{0.2748}$,
close to the scaling $\Nuss \sim \Ray^{2/7}$, found in similar conditions~\cite{johnston_Comparison_2009}. 
Interestingly, this range of Rayleigh numbers produces an approximately constant heat transport scaling, in spite of the transitions of increasing complexity of the flow.
This is consistent with Grossmann-Lohse theory, which states that the ratio between dissipation in the boundary layer and the bulk determines the heat transport dynamics. We verified (calculated but not shown) that this ratio does not change in the range of Rayleigh numbers studied.


Comparing the results between the DNS and the invariant solutions found,
the steady state (ST) sustains the highest Nusselt number across the range of $\Ray$ studied, consistent with previous findings~\cite{wen_steady_2022}. The periodic orbits and the DNS fall below. 
The PO1 branch initially sustains the heat transport observed in the DNS, but at higher $\Ray$ yields a consistently smaller transport efficiency.
The PO2 matches very closely the DNS in the range where it is present. On the other hand, the heat transport of PO3 is almost independent of $\Ray$, branching off from ST at $\Ray=\num{8e6}$.
These differences indicate that the invariant solutions do not trace the mean heat transport of the DNS. Instead, they represent distinct dynamical states with different transport efficiencies.

\REVB{The energy shown in \fref{fig:nusselt-energy}(b) exhibits an ordering of the branches roughly reversed with respect to the heat transport. 
ST, which sustains the highest $\Nuss$, carries the lowest energy of all solutions once it becomes unstable, indicating that its steady roll structure is the most efficient at transporting heat per unit of energy. Conversely, PO1 follows the energy of the DNS closely at first, but sustains a consistently higher energy for $\Ray\gtrsim \num{4e6}$ despite its lower heat transport.
PO2 once again closely matches the DNS in the range where it is present. The energy of PO3 in contrast to its almost constant heat transport grows steeply after branching off from ST.}

\section{Conclusions}
\label{sec:conclusions}

We have computed and continued, via a Newton-Krylov-Hookstep method coupled to direct numerical simulations, a steady state and three families of periodic orbits of a 2D RB flow at $\Pr=1$ across nearly two decades of Rayleigh number, $10^5<\Ray<\num{2e7}$. Floquet analysis of each branch identified the bifurcation type at every change of stability and provided the spatial structure of the most unstable perturbations. 
The Floquet modes consistently localize on the thermal plumes, serving as the primary source of instability throughout the route to chaos. We show that in this route the flow visits several orbits, even though the main frequency of the flow remains approximately the same. We determined too that the onset of chaos, signaled by a positive leading Lyapunov exponent at $\Ray \approx \num{1.4e7}$, is preceded by an $S_2$ symmetry-breaking event at $\Ray\approx\num{9e6}$. This implies that, in this geometry, chaos emerges from gradual amplification of instabilities within an already asymmetric state, rather than as a direct consequence of the symmetry break.
Heat transport measurements show that the steady state sustains the highest Nusselt number throughout the explored range, and that bifurcated periodic orbits track the DNS heat transport only within the Rayleigh-number windows where they are dynamically relevant. Despite the successive dynamical transitions, the overall scaling remains robust, with $\Nuss \sim \Ray^{0.27}$ holding across all studied regimes.

Several directions remain open.
Although analysis of the flow strongly hints at the presence of unstable periodic orbits between $\Ray\approx \num{6e6}$ and the onset of chaos at $\Ray \approx \num{1.4e7}$, we failed to find any. We believe this is a limitation of the Newton-Krylov method used which struggles with longer orbits. In future work we will implement new techniques to obtain invariant solutions that do not rely on a shooting strategy. Furthermore, we will explore the relationship between the onset of chaos, the loss of synchronization properties \cite{agasthya_Reconstructing_2022} and the sharp increase in dimensionality \cite{Vinograd_Clark}.
The same framework should be extended to three-dimensional Rayleigh-Bénard convection, where additional symmetries and a richer set of coherent structures are expected to organize the dynamics.

\begin{acknowledgments}
The authors thank José Eduardo Wesfreid for helpful feedback during the development of this work. JC and MYV are supported by the UdeSA PhD Fellowship program, MYV is also supported by the Google PhD Fellowship program.
\end{acknowledgments}

Code is available at \url{https://github.com/joacocullen/UPOsinRB}


\bibliography{bib}

\end{document}